\documentclass{llncs}
\usepackage{amssymb}
\usepackage{fancyvrb}
\usepackage{rotating}
\usepackage{paralist}
\usepackage{colortbl}
\usepackage{algorithm}
\usepackage{algorithmic}
\usepackage{epsfig}
\usepackage{graphicx}
\usepackage{color}
\usepackage{multirow}
\usepackage{tabularx}
\usepackage{epstopdf}
\usepackage{url}

\usepackage{subcaption}
\usepackage{amsmath}
\usepackage[compatibility=false]{caption}
\DeclareCaptionType{copyrightbox}
\usepackage{subcaption}
\usepackage[numbers]{natbib}
\usepackage{float}

\usepackage{adjustbox}

\DeclareMathOperator*{\argmin}{arg\,min}

\newcolumntype{H}{>{\columncolor{black}\color{white}}c}

\graphicspath{{./figures/}}

\begin{document}

\title{Modifying iterated Laplace approximations}

\author{Tiep Mai\inst{1} \and Simon Wilson\inst{2}}
\institute{Bell Laboratories, Ireland\\
\url{tiep.mai@alcatel-lucent.com}
\and 
School of Computer Science and Statistics, Trinity College, Dublin 2, Ireland\\
\url{swilson@tcd.ie}}

\maketitle


\begin{abstract}
In this paper, several modifications are introduced to the functional approximation method iterLap \citep{ref:Bornkamp2011} to reduce the approximation error, including stopping rule adjustment, proposal of new residual function, starting point selection for numerical optimisation, scaling of Hessian matrix. Illustrative examples are also provided to show the trade-off between running time and accuracy of the original and modified methods. 
\end{abstract}


\section{Introduction}
\label{sec:intro}
Compared to Monte Carlo methods, functional approximation does not have the asymptotic convergence property, e.g. even when a method is applied repeatedly, the approximation may not converge to the target density at all. Still, these methods are extremely useful in a practical application where time is the most invaluable resource. Furthermore, functional approximations can be used to complement Monte Carlo methods (used as importance sampling or MCMC proposal functions). For an IS algorithm, this may produce samples appropriately in the target support and result in evenly weighted samples. A MCMC trajectory may move adaptively to the local correlation or escape a modal trap by using these approximations.

Among all methods, Laplace approximation \citep{ref:Laplace1774} is the simplest but still very useful. In general, this method approximates a target non-normalised density by a normal distribution. Laplace approximation has all the useful properties of a normal distribution, e.g. the normal form of the conditional and marginal distributions.

As distributions tend to converge to normal form asymptotically, Laplace approximation is very efficient in these cases. However, for non-normal distribution, this approximation suffers from two shortcomings. Firstly, it only works well with a uni-modal function and ignores the other modes if the target density is multi-modal. Secondly, the normal distribution implies a linear correlation between random variables and hence cannot accommodate non-linear dependency. 

Variational Bayes \citep{ref:Jordan1999} and expectation propagation \citep{ref:Bishop2006} are alternative solutions. However, as these methods aim to minimise the Kullback-Leibler (KL) divergence over an approximated density of specific form, they are only efficient when the KL divergence can be evaluated in closed form. Furthermore, these solutions are not as plugin as Laplace approximation due to the problem-specific derivation of KL divergence.

As an extension of the normal form in Laplace approximation, the mixture of normal distributions is a very flexible family of distributions, capable of  accommodating multi-modal and non-linear functions. A numerical estimation procedure for the weights, means and variances of the mixture form was given in \citep{ref:Gelman2003}. Still, there are issues with this algorithm. With a skewed uni-modal target density, all the modes and precisions, will be almost same (the difference is caused by numerical computation) and the resulting approximation is symmetric (by normal distribution), not being able to reflect the skewness of the target. Furthermore, it may be wasteful to do multiple simultaneous mode optimisation by random starting points. 

Addressing these issues, Bornkamp \citep{ref:Bornkamp2011} proposed the iterated Laplace approximation (iterLap), correcting the approximation discrepancy iteratively by adding a new component at an appropriate location.

Another way to approximate a target density by a Gaussian mixture is to use expectation maximisation (EM) algorithm \citep{ref:Bishop2006}. Unlike above methods, the EM solution relies on samples of the target density instead of numerical evaluation. More complex solutions are also needed to address the issue of unknown number of Gaussian components \citep{ref:Verbeek2003}.

In this work, we propose a modified version of iterLap which is a direct approximation to numerical evaluation of a target density. It is shown by various experiments that this modified version is slower but provides smaller approximation error, compared to the original iterLap solution. 

So, Section \ref{sec:iterLap} provides a description of original iterLap algorithm. Then, Section \ref{sec:miterLap} gives an overview and discussion over several modifications of a new approximation. Next, we show various numerical examples to compare the trade-off of running time and approximation error of two algorithms in Section \ref{sec:comp}. Finally, Section \ref{sec:conclusion} concludes this work.


\section{Iterated Laplace approximation (iterLap)}
\label{sec:iterLap}

The description of iterLap algorithm is given in Algorithm \ref{alg:iterLap} \citep{ref:Bornkamp2011,ref:Bornkamp2011B}.

\begin{algorithm}[!ht]
\caption{iterLap}
\label{alg:iterLap}
\begin{enumerate}
\item Find $n_1$ local minima $\mu_i$ and their corresponding Hessian matrices $Q_i$ of $g(\cdot)=-\log(q_x(\cdot))$ (like the above algorithm). Let $n_c$ be the current number of components (initially, $n_c=n_1$). The current approximation $\widetilde{q}_{n_c;x}(\cdot)$ with unknown non-normalised weights is:
\begin{align}
\label{sec_iterLap:iterlap_proc2_mixture_approximation}
\widetilde{q}_{n_c;x}(x) &= \sum_{i=1}^{n_c} w_i N(x|\mu_i,Q_i).
\end{align}
\item Assume that for each component $i$, there are $n_x$ location vectors $x_{g;i,j}$ ($j=1:n_x$) generated from the distribution $N(\mu_i,Q_i)$; $d_x$ is the length of vector $x$. Let $X$ be a $m \times d_x$ matrix comprising all the location vectors $x_{g;i,j}$ and the mean $\mu_i$ (each row $X_{k,1:d_x}$ ($k=1:m$) is either $x_{g;i,j}$ or $\mu_i$). Let $y$ be a vector of length $m$ comprising the values of the target density, evaluated for all locations in $X$: $y_k=q_{x}(X_{k,1:d_x})$. Let $Z$ be a $m \times n_c$ matrix, satisfying: $Z_{k,i} = N(X_{k,1:d_x}|\mu_i,Q_i)$. The weights $w=w_{1:n_c}$ can be obtained by using quadratic programming: 
\begin{align}
\label{sec_iterLap:iterlap_proc2_quadratic_programming}
w &= \argmin_{b} [ (y-Zb)^{T}(y-Zb) ], ~~~ b \geq 0.
\end{align}
\item Define a residual function $g_{n_c}(\cdot)$ with:
\begin{align}
\label{sec_iterLap:iterlap_proc2_residual_function}
z &= q_{x}(x)-\widetilde{q}_{n_c;x}(x),\\
g_{n_c}(x) &= \begin{cases}
      -\log(z) ~&\text{if}~ z \geq z_{l} \\
      -\log( \exp(z-z_{l}) z_{l} ) ~&\text{if}~ z < z_{l}
    \end{cases}.
\end{align}
A lower bound $z_{l}=1e^{-4}$ is used in the R package iterLap \citep{ref:Bornkamp2011B}. Find a new component by minimising the residual function:
\begin{align}
\label{sec_iterLap:iterlap_proc2_new_component}
\mu_{n_c+1} &= \argmin_x \{g_{n_c}(x)\},\\
Q_{n_c+1} &= \left. \frac{\partial^2 g_{n_c}(x)}{\partial x^2} \right|_{x=\mu_{n_c+1}}.
\end{align}
The starting points of this optimisation step are chosen by checking the ratio $q_{x}(x) / \widetilde{q}_{n_c;x}(x)$. Points with large ratios are preferred (the locations where $\widetilde{q}_{n_c;x}(x)$ underestimates $q_{x}(x)$). 
\item Update the number of components: $n_c = n_c + 1$. If one of the following criteria is satisfied:
  \begin{itemize}
  \item $n_c \geq n_{c;max}$ where $n_{c;max}$ is a predefined maximum number of components.
  \item $|q_{x}(x)-\widetilde{q}_{n_c;x}(x)| \leq \delta$ with a predefined error bound $\delta$.
  \item $\sum_{i=1}^{n_c} w_i$ does not improve (by comparing with previous sums of weights).
  \item Cannot find a new component.
  \end{itemize}
then re-estimate the weights by step $2$ and stop the algorithm. Otherwise, repeat steps $2 \rightarrow 4$.
\end{enumerate}
\end{algorithm}

\begin{figure}[!ht]
\centering
\begin{adjustbox}{center}
\begin{subfigure}[b]{0.55\textwidth}
    \centering
    \includegraphics[scale=0.6]{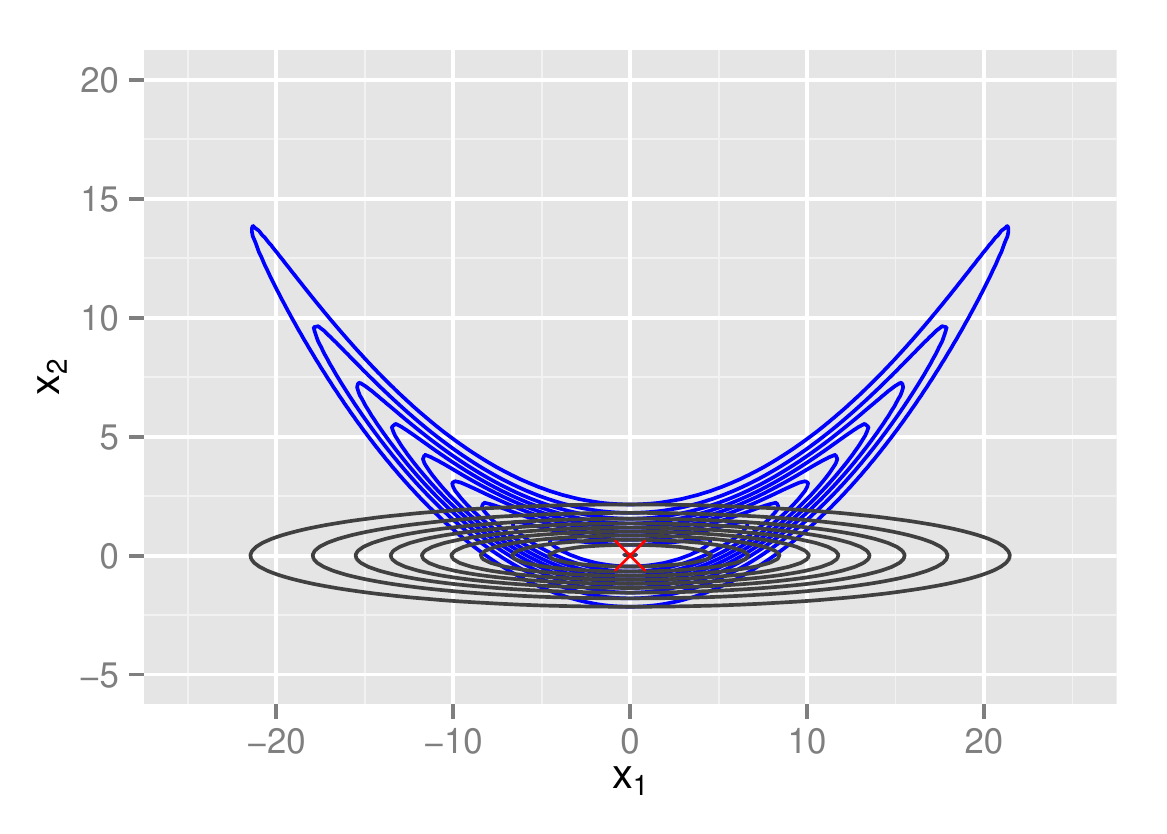}
    \subcaption{Laplace approximation}
    \label{sec_iterLap_fig:iterLap_ex01_a_maxDim01}
\end{subfigure}
\begin{subfigure}[b]{0.55\textwidth}
    \centering
    \includegraphics[scale=0.6]{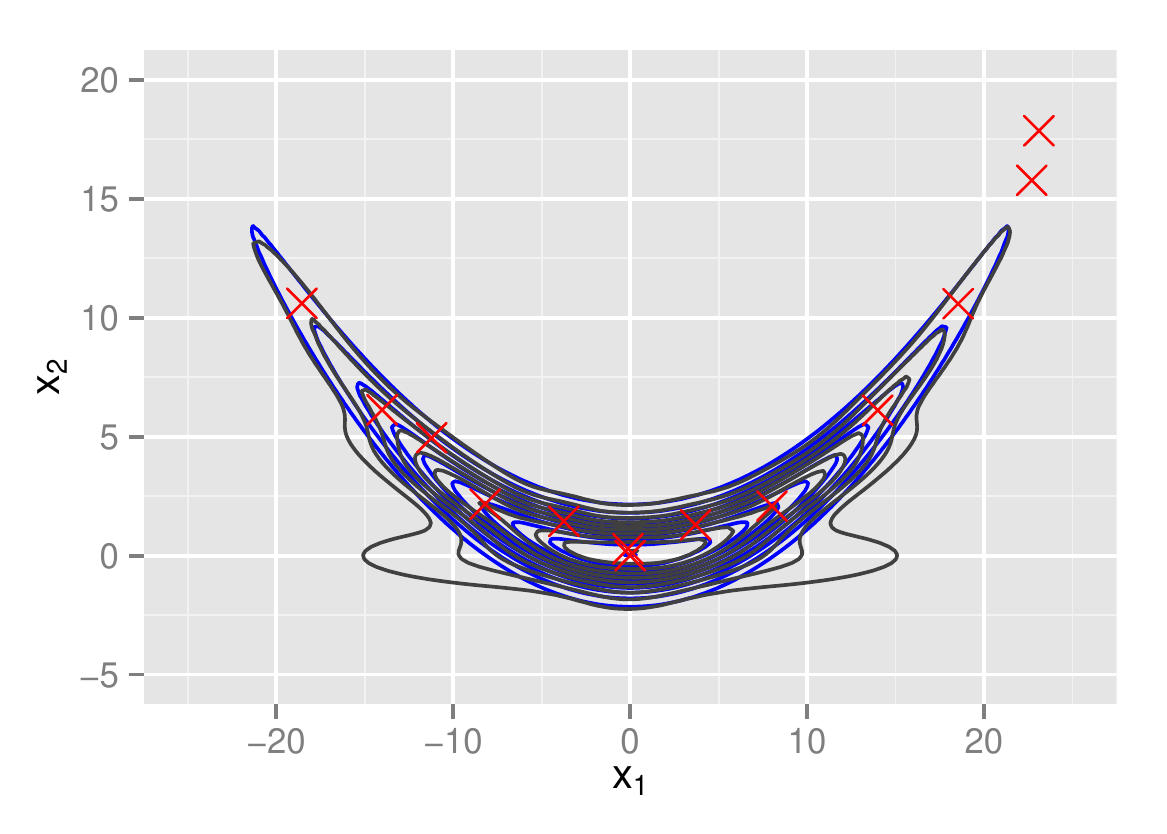}
    \subcaption{iterLap approximation}
    \label{sec_iterLap_fig:iterLap_ex01_a_maxDim50}
\end{subfigure}
\end{adjustbox}

\caption{Laplace and iterLap approximations. The target density and approximations are drawn by blue and black contours accordingly. The red crosses are the means of normal components. The iterLap approximation has $15$ components.} 
\label{sec_iterLap_fig:iterLap_ex01}
\end{figure}

To illustrate the performance difference between Laplace and iterLap approximations, Figure \ref{sec_iterLap_fig:iterLap_ex01} shows both approximations for the following target density of a random variable $x=(x_1,x_2)$: 
\begin{align}
\label{sec_iterLap:iterlap_example_target}
q_{x}(x) \propto N(x_1|0,\sigma_{1}^2=10^2) N(x_2|\mu_2 = 0.03 x_1^2,\sigma_{2}^2=1^2).
\end{align}
In Figure \ref{sec_iterLap_fig:iterLap_ex01}, iterLap approximation is much better in terms of adapting to the non-linear dependency between two variables. Still, the iterLap method can be further improved by some modifications, which will be discussed later.

\section{Modifying the iterLap method}
\label{sec:miterLap}

Based on Section \ref{sec:iterLap}, in this section, we propose several modifications to improve the performance of iterLap approximation.

\vspace{+2em}
\noindent \textit{Stopping rule by the normalising constant}

\noindent As a reminder, iterLap approximates a non-normalised density $q_{x}(\cdot)$ at iteration $n_c$ by: 

\begin{align}
\label{sec_miterLap_eqn:iterLap2_approx}
q_{x}(x)&\approx \widetilde{q}_{n_c;x}(x) = \sum_{i=1}^{n_c} w_i N(x|\mu_i,Q_i),
\end{align}
where $\mu_i$, $Q_i$ are found by an optimisation procedure and non-normalised weights $w_i$ are estimated by quadratic programming in Section \ref{sec:iterLap}. By the above approximation, the normalising constant $\zeta$ of $q_{x}(\cdot)$ is estimated by:
\begin{align}
\label{sec_miterLap_eqn:iterLap2_normconst}
\zeta &= \int q_{x}(x)dx \approx \widetilde{\zeta}_{n_c} = \int \widetilde{q}_{n_c;x}(x)dx = \sum_{i=1}^{n_c} w_i.
\end{align}
The constant $\widetilde{\zeta}_{n_c}$ represents the probability mass of the approximated density $\widetilde{q}_{n_c;x}(x)$. Hence, in \citep{ref:Bornkamp2011}, the iterative process stops when $\widetilde{\zeta}_{n_c}$ does not improve any more, i.e. satisfying the following equation:
\begin{align}
\label{sec_miterLap_eqn:iterLap2_normconst_stoprule}
|\widetilde{\zeta}_{n_c}-0.5(\widetilde{\zeta}_{n_c-1}+\widetilde{\zeta}_{n_c-2})|/\widetilde{\zeta}_{n_c} &< \delta_{\zeta},
\end{align}
with a predefined threshold $\delta_{\zeta}$.

However, even though a new component $N(x|\mu_i,Q_i)$ does not improve the estimated volume of $\widetilde{q}_{n_c;x}(x)$, it may still correct the density and decrease the approximation error. Furthermore, a new component generates more explored points which may be useful in the optimisation and quadratic programming step. Hence, we remove this stopping criterion from the iterLap source code.

\vspace{+2em}
\noindent \textit{Residual function}

\noindent At each iteration, iterLap \citep{ref:Bornkamp2011B} finds a new component by minimising a residual function $g_{a;n_c}(x)$ with:
\begin{align}
\label{sec_miterLap_eqn:iterLap2_resfunc_g_a}
z &= q_{x}(x)-\widetilde{q}_{n_c;x}(x),\\
g_{a;n_c}(x) &= \begin{cases}
      -\log(z) ~&\text{if}~ z \geq z_{l} > 0 \\
      -\log( \exp(z-z_{l}) z_{l} ) ~&\text{if}~ z < z_{l}
    \end{cases},
\end{align}
where $z_{l}$ is a predefined positive lower bound. The new component's mean and precision matrix are obtained by:
\begin{align}
\label{sec_miterLap_eqn:iterlap2_findnewcomp_meanprecision}
\mu_{n_c+1} &= \argmin_x \{g_{a;n_c}(x)\},\\
Q_{n_c+1} &= \left. \frac{\partial^2 g_{a;n_c}(x)}{\partial x^2} \right|_{x=\mu_{n_c+1}}.
\end{align}

We will use the following example to illustrate and discuss the above step.

\begin{example}
\label{sec_miterLap_exa:iterLap2_exa05}
Consider a non-normalised density $q_{x}(\cdot)$:
\begin{align}
\label{sec_miterLap_eqn:iterLap2_exa05_targetq}
q_{x}(x) &= \exp{\left( -\frac{1}{2 \times 5^2}x^2  -\frac{1}{2 \times 5^2}(|x|-0.5)_{+}^3 \right)},
\end{align}
where $a_{+} = a~\text{if}~a \geq 0$ and $a_{+} = 0~\text{if}~a < 0$.
\end{example}

\begin{figure}[!ht]
\centering
\includegraphics[scale=0.6]{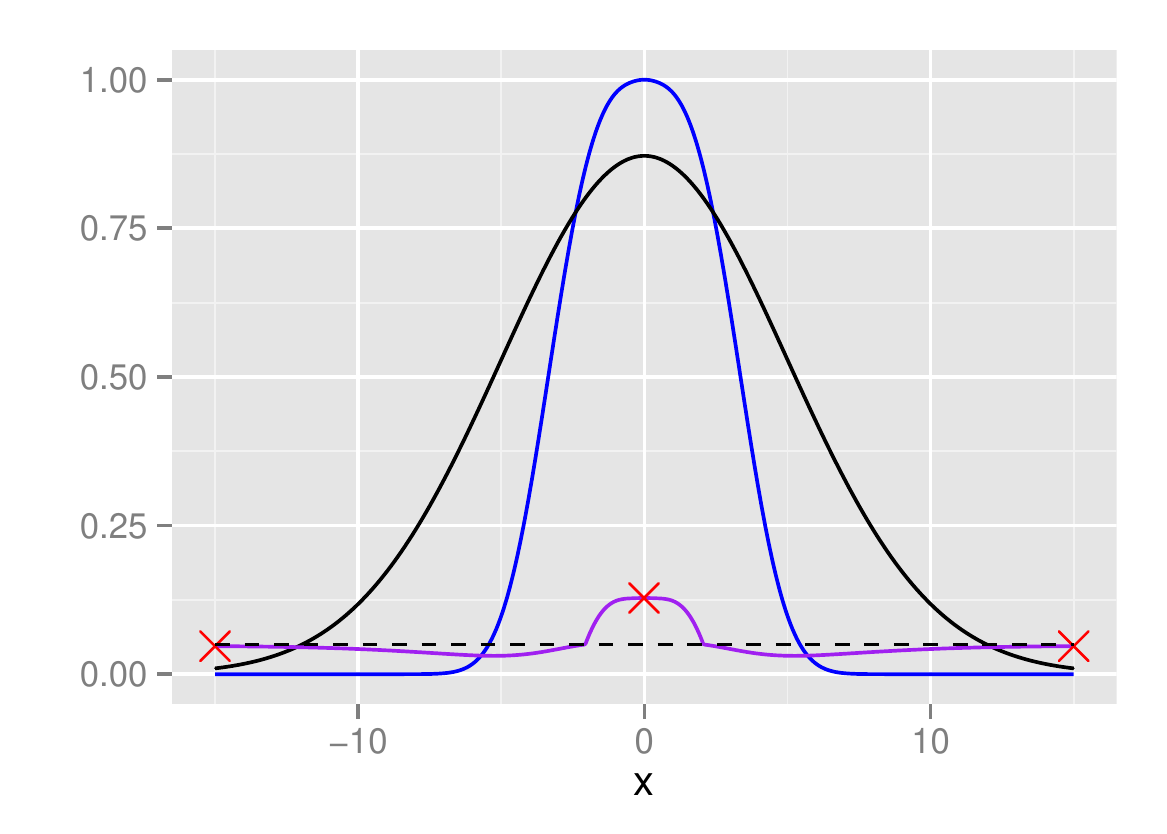}
\caption{Example \ref{sec_miterLap_exa:iterLap2_exa05}: the blue and black lines show the target density $q_{x}(\cdot)$ and the approximated density $\widetilde{q}_{1;x}(\cdot)$; the purple line is $\exp{(-g_{a;1}(\cdot))}$ with red crosses as maximum points (the leftmost and rightmost crosses are actually $\inf$ and $-\inf$); the lower bound $z_{l}=0.05$ is marked by dashed horizontal line.}
\label{sec_miterLap_501_fig:ex05_residual_function_g_a}
\end{figure}

At iteration $1$, $q_{x}(\cdot)$ is approximated by $\widetilde{q}_{1;x}(\cdot)$ with a single normal component and the residual function is $g_{a;1}(\cdot)$. The functions $q_{x}(\cdot)$, $\widetilde{q}_{1;x}(\cdot)$ and $\exp{(-g_{a;1}(\cdot))}$ are plotted in Figure \ref{sec_miterLap_501_fig:ex05_residual_function_g_a}. The maximum points of $\exp{(-g_{a;1}(\cdot))}$ or equivalently, the minimum points of $g_{a;1}(\cdot)$, are also shown, representing the next potential component means. 

Figure \ref{sec_miterLap_501_fig:ex05_residual_function_g_a} shows that there are two potential maximum points at $\inf$ and $-\inf$, which are not good locations for new components at all because they do not have any effect on the approximation. So, these locations should be avoided to save computation time. 

From Equation \ref{sec_miterLap_eqn:iterLap2_resfunc_g_a} and Figure \ref{sec_miterLap_501_fig:ex05_residual_function_g_a}, it can be seen that iterLap prefers choosing the locations at which $q_{x}(x)$ is significantly underestimated by $\widetilde{q}_{n_c;x}(x))$. The locations in the overestimated region ($q_{x}(x) <  \widetilde{q}_{n_c;x}(x))$) are ignored. However, by experimenting with several residual functions and examples, we find that adding new components in the overestimated region does improve the approximation. Partially, this may be because the explored variable space is extended by a more lenient rule, which in turn improves the optimisation and quadratic programming steps. 

\begin{figure}[!ht]
\centering
\begin{adjustbox}{center}
\begin{subfigure}[b]{0.55\textwidth}
    \centering
    \includegraphics[scale=0.6]{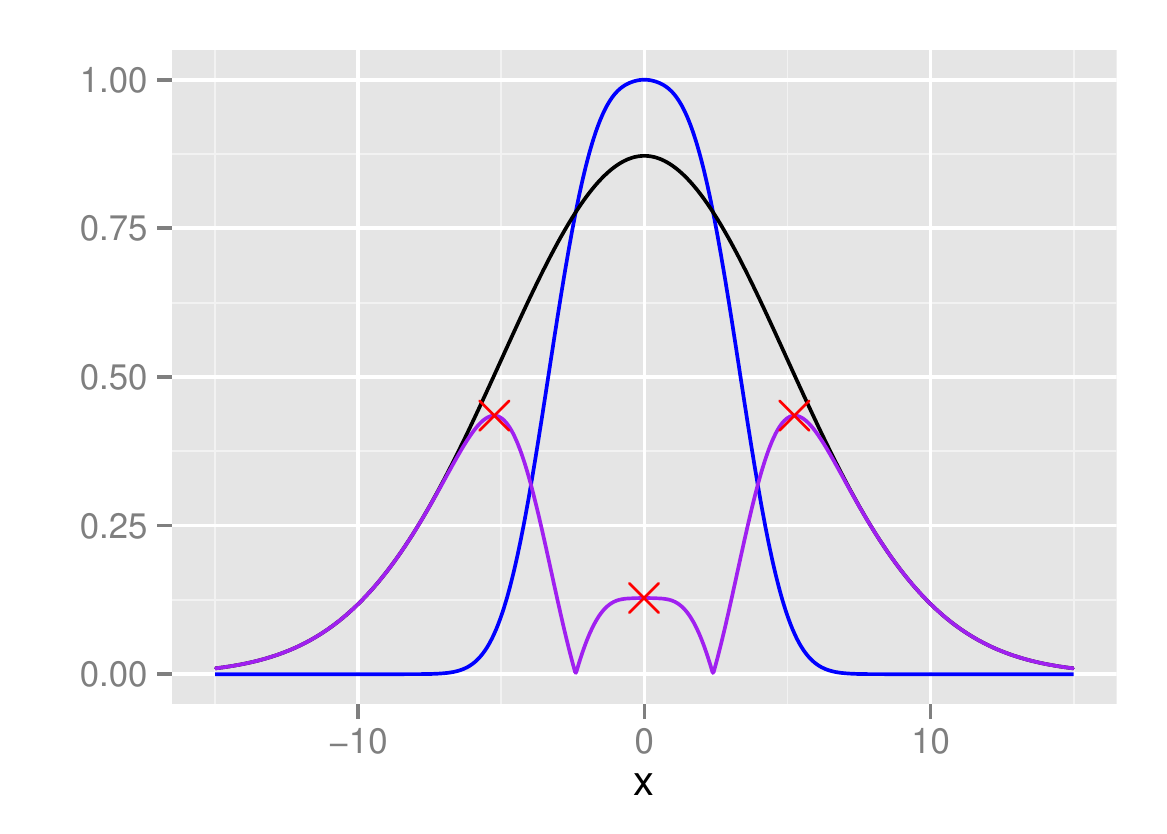}
    \subcaption{$\alpha=0$}
    \label{sec_miterLap_502_fig:ex05_residual_function_g_b_coefa1}
\end{subfigure}
\begin{subfigure}[b]{0.55\textwidth}
    \centering
    \includegraphics[scale=0.6]{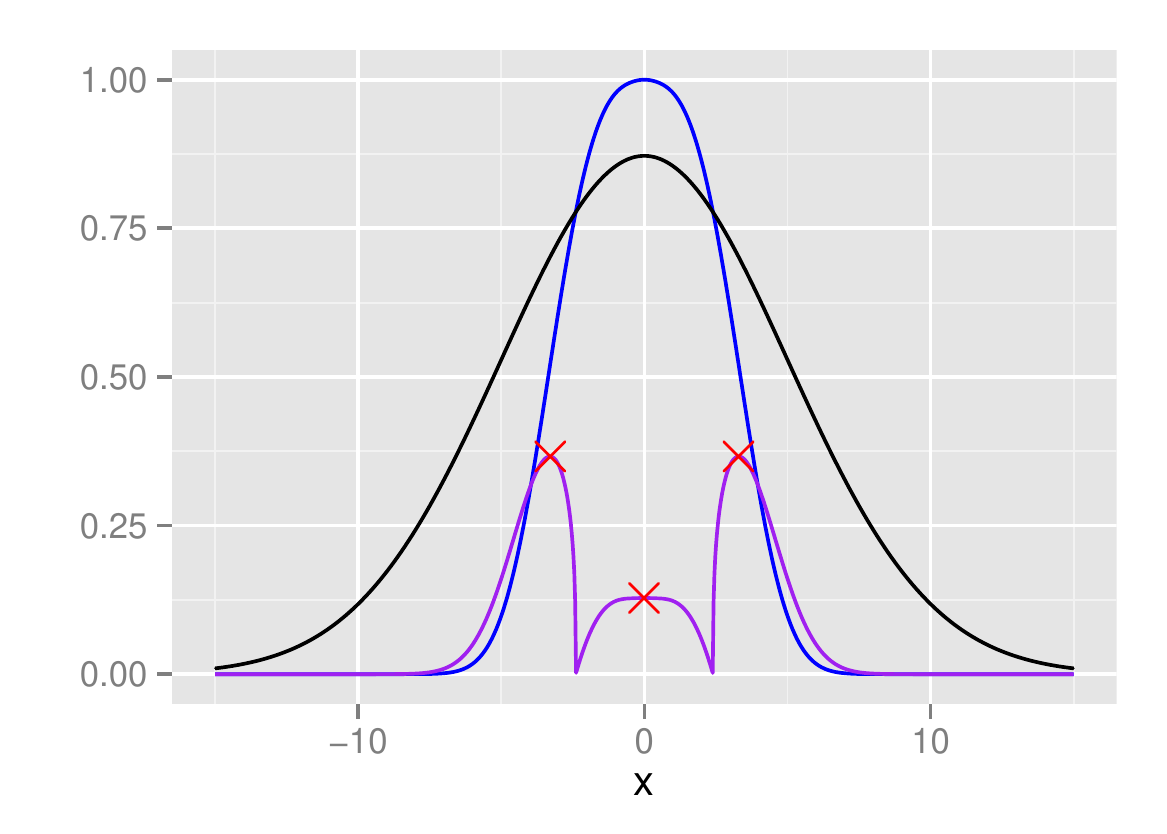}
    \subcaption{$\alpha=3$}
    \label{sec_miterLap_502_fig:ex05_residual_function_g_b_coefa3}
\end{subfigure}
\end{adjustbox}

\caption{Example \ref{sec_miterLap_exa:iterLap2_exa05}: the blue, black and purple lines show the target density $q_{x}(\cdot)$, the approximated density $\widetilde{q}_{1;x}(\cdot)$ and the function $\exp{(-g_{b;1}(\cdot))}$ with red crosses as maximum points of $\exp{(-g_{b;1}(\cdot))}$.} 
\label{sec_miterLap_502_fig:ex05_residual_function_g_b}
\end{figure}

So, we propose a new residual function $g_{n_c;r_b}(x)$ with:
\begin{align}
\label{sec_miterLap_eqn:iterLap2_resfunc_g_b}
z &= q_{x}(x)-\widetilde{q}_{n_c;x}(x),\\
g_{b;n_c}(x) &= \begin{cases}
      -\log(z + \epsilon_z) ~&\text{if}~ z \geq 0 \\
      -[\log(-z + \epsilon_z) + \alpha(\log{(q_{x}(x))} - lq_{max;x}) ]/(1+\alpha) ~&\text{if}~ z < 0
    \end{cases},
\end{align}
where $lq_{max;x}$ is the maximum log value of $\log{(q_{x}(x))}$ until the current iteration; a very small constant $\epsilon_z=e^{-10}$ is used only for the positivity condition of the log operator; $\alpha>0$ is an optional coefficient which pulls the optimisation point of $g_{b;n_c}(x)$ to a location of high log density value, $\log{(q_{x}(x))}$. Figure \ref{sec_miterLap_502_fig:ex05_residual_function_g_b} illustrates the above residual function $g_{n_c;r_b}(x)$ with $\alpha=0$ and $\alpha=3$.


\vspace{+2em}
\noindent \textit{Select starting points for the optimisation}

\noindent Using different starting points for the optimisation of $g_{b;n_c}(x)$ results in different component means. Bornkamp \citep{ref:Bornkamp2011} uses the ratio $q_{x}(x)/\widetilde{q}_{n_c;x}(x)$ as the criterion of choosing the optimisation starting points. However, such a criterion may lead to a point located at the distribution tail. For example, with a t-distribution $q_{x}(x)$ and a normal distribution $\widetilde{q}_{n_c;x}(x)$, the ratio is extremely large at the tail. 

So, we use the absolute difference $|q_{x}(x) - \widetilde{q}_{n_c;x}(x)|$ as a selection criterion which is closely related to the residual function $g_{b;n_c}(x)$. As a reminder of Section \ref{sec:iterLap}, iterLap keeps a $m \times d_x$ matrix $X$ of all explored locations, a vector $y$ of target density values $q_{x}(\cdot)$ and a $m \times n_c$ matrix $Z$ of component density values $N(\cdot|\mu_i,Q_i)$ evaluated at the explored locations. The algorithm to select starting points $X_s$ from $X$ is shown in Algorithm \ref{alg:iterLap_sp}. 

\begin{algorithm}[ht]
\caption{Select optimisation starting points}
\label{alg:iterLap_sp}
\begin{enumerate}
\item Let $X_a=X$. Remove from $X_a$ points $x_i$ with small target density values, i.e. satisfying the condition: $\log{(q_{x}(x_i))}-lq_{max;x} < \delta_{lq}$
\item Find a point $x_j$ in $X_a$ which maximises $|q_{x}(x_j) - \widetilde{q}_{n_c;x}(x_j)|$. Add that point $x_j$ to $X_s$.
\item Remove from $X_a$ points $x_k$ close to $x_j$.
\item Stop the algorithm if there are enough starting points in $X_s$ or if $X_a$ is empty. Otherwise, repeat steps $2 \rightarrow 4$.
\end{enumerate}
\end{algorithm}

\vspace{+2em}
\noindent \textit{A note on standardising $q_{x}(\cdot)$ and $\widetilde{q}_{x}(\cdot)$}

\noindent Usually, the comparison between two densities is done based on their normalised densities. However, aside from the difficulty of obtaining the normalising constant, there is another issue. It is almost impossible to have a "good" approximation at the tail by any method. The absolute difference $|q_{x}(x) - \widetilde{q}_{x}(\cdot)|$ at a specific location at the tail may be extremely small but in high dimension space, the sum of all the absolute differences in every direction of the tail may become significant. Consequently, even though $\widetilde{q}_{x}(\cdot)$ is a "good" approximation of $q_{x}(\cdot)$ locally, the normalised $\widetilde{p}_{x}(\cdot)$ is however a "poor" approximation of the normalised $p_{x}(\cdot)$. 


Hence, instead of globally normalising $q_{x}(\cdot)$ and $\widetilde{q}_{x}(\cdot)$, we only standardise these two densities on a user-defined grid $X_g$. $q_{x}(\cdot)$ is evaluated for all grid points $x_j$ and the standardised density $r_{x}(\cdot)$ on the grid is obtained by:
\begin{align}
\label{sec_miterLap_eqn:iterlap2_standardised_density}
r_{x}(x_j) &= \frac{q_{x}(x_j)}{\sum_{j^{\prime}}q_{x}(x_{j^{\prime}})}.
\end{align}
The standardised $\widetilde{r}_{x}(\cdot)$ is obtained in a similar manner. So, local comparison of two densities can be done by analysing the standardised densities. There is one statistic $s(r_{x},\widetilde{r}_{x})$ that we find reasonable for the comparison purposes between two standardised densities:
\begin{align}
\label{sec_miterLap_eqn:iterlap2_standardised_density_compstat}
s(r_{x},\widetilde{r}_{x}) &= \sum_{x_i \in X_g} |r_{x}(x_i) - \widetilde{r}_{x}(x_i)|.
\end{align}
With the same grid $X_g$, $s(r_{x},\widetilde{r}_{a;x})$ can be compared with another $s(r_{x},\widetilde{r}_{b;x})$.

\vspace{+2em}
\noindent \textit{Scaling a component's Hessian}

\noindent In iterLap, the Hessian of $g_{b;n_c}(x)$ at the mode is used as the component's precision matrix. Usually, the sharper the curvature of $g_{b;n_c}(x)$, the larger the eigenvalues of the Hessian matrix at the mode and the more focused the corresponding normal component. 

In some cases, the numerically evaluated Hessian matrix at the mode is sharper than the actual curvature and iterLap may fail to improve such a target density. In Example \ref{sec_miterLap_exa:iterLap2_exa05}, the function $\log{(-q_{x}(x) )}$ is a quadratic function near the mode $x=0$ but it becomes a cubic function with much sharper curvature when $|x|>0.5$. As a result, the approximated $\widetilde{q}_{1;x}(x)$ at the first iteration is much flatter than the target density, which is shown in Figure \ref{sec_miterLap_502_fig:ex05_residual_function_g_b}. Unfortunately, in this example, even with more extra components, the iterLap approximation does not improve.

There are two ways to get around this issue. Firstly, we can allow a manual scaling of the Hessian matrix. With a user-defined scale factor $\kappa_a$, a new component is added with a precision matrix:
\begin{align}
\label{sec_miterLap_eqn:iterlap2_scalehess_01}
Q_{n_c+1} &= \kappa_a \left. \frac{\partial^2 g_{a;n_c}(x)}{\partial x^2} \right|_{x=\mu_{n_c+1}}.
\end{align}
This modification is analogous to using kernel density estimation with a small bandwidth. A component with large precision matrix affects the approximation in a small region. 

\begin{figure}[!ht]
\centering
\begin{adjustbox}{center}
\begin{subfigure}[b]{0.55\textwidth}
    \centering
    \includegraphics[scale=0.6]{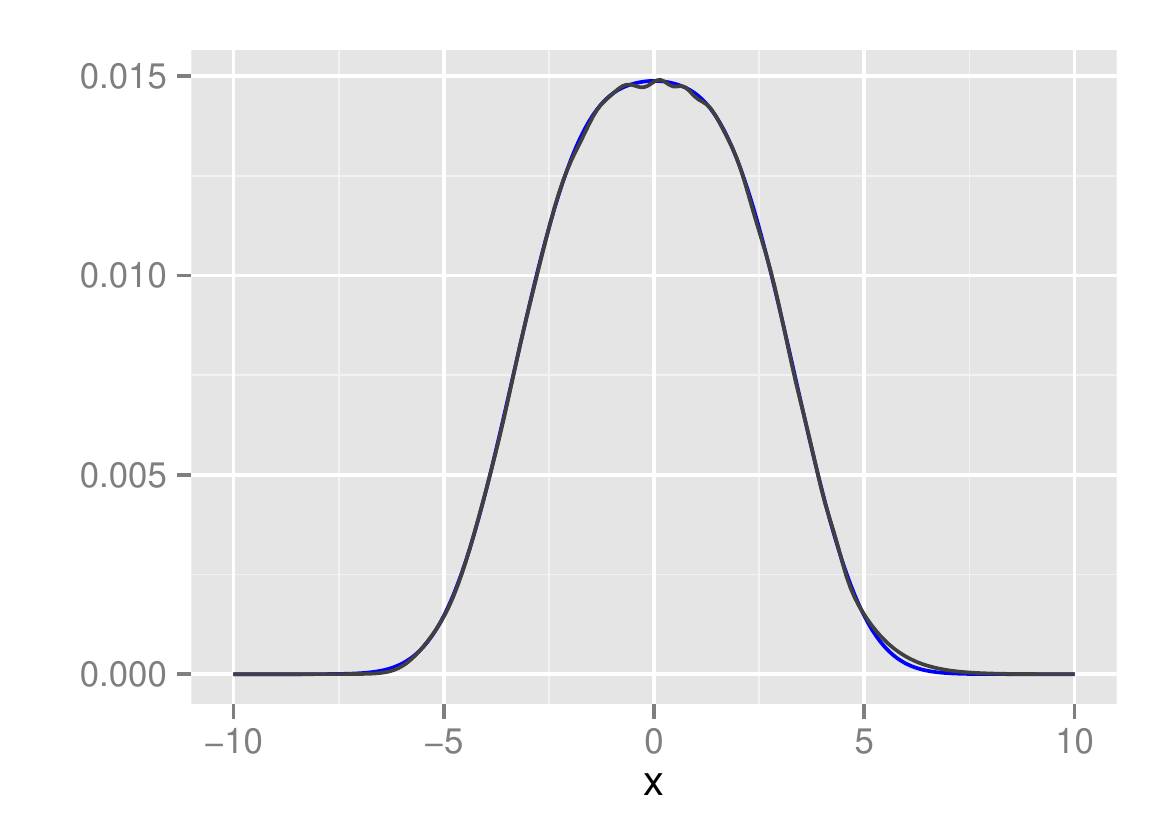}
    \subcaption{$\kappa_a=1.5$ with no multiplicative scaling. $\widetilde{r}_{n_c;x}(\cdot)$ has $n_c=14$ components}
    \label{sec_miterLap_503_fig:ex05_approx01_hessianscale15}
\end{subfigure}
\begin{subfigure}[b]{0.55\textwidth}
    \centering
    \includegraphics[scale=0.6]{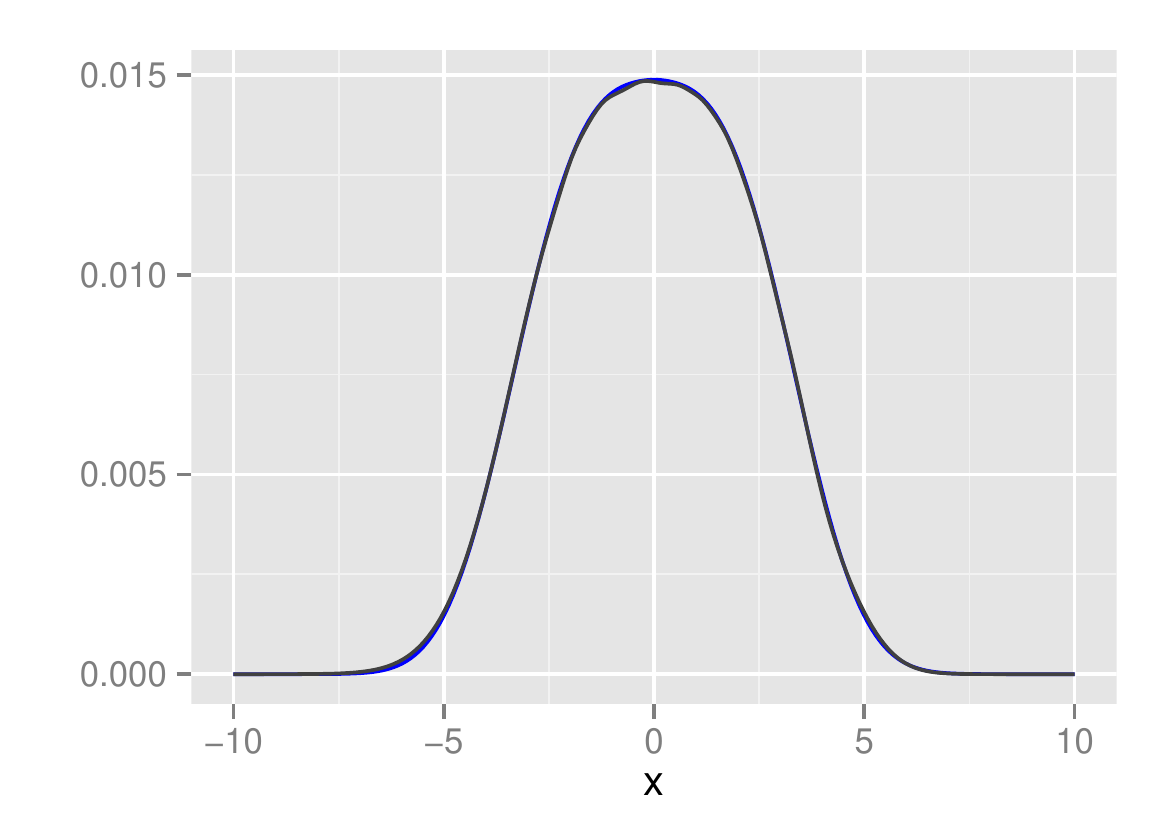}
    \subcaption{Multiplicative scaling with $n_{dup}=3$ and $\kappa_b=1.25$. $\kappa_a=1$. $\widetilde{r}_{n_c;x}(\cdot)$ has $n_c=12$ components}
    \label{sec_miterLap_503_fig:ex05_approx02_dupmean_hessian_scale125}
\end{subfigure}
\end{adjustbox}

\caption{Example \ref{sec_miterLap_exa:iterLap2_exa05}: the iterLap approximation with hessian scaling and multiplicative scaling; the blue and black curves are the standardised densities $r_{x}(\cdot)$ and $\widetilde{r}_{n_c;x}(\cdot)$.}
\label{sec_miterLap_503_fig:ex05_approx}
\end{figure}

With $\kappa_a=1.5$, the standardised densities for Example \ref{sec_miterLap_exa:iterLap2_exa05} are shown in Figure \ref{sec_miterLap_503_fig:ex05_approx01_hessianscale15}. Clearly, the approximation becomes much better.

One potential problem with this manual scaling is that all precision matrices become larger and all corresponding components are sharper, affecting only small local regions centred at the means. Similar to the case of using a small bandwidth in kernel method, the approximation becomes wriggly and only gets smoother with a larger number of components. 

In a second attempt, we try to only adjust the Hessian matrix when needed. In Figure \ref{sec_miterLap_502_fig:ex05_residual_function_g_b}, one of the potential means for the next components is $\mu=0$ (by the optimisation of $g_{b;1}(\cdot)$). As $\mu=0$ is already a component mean of the approximated  $\widetilde{q}_{1;x}(\cdot)$, iterLap will not accept $\mu=0$ as a mean of a new component and try to find other locations. Also, even if $\mu=0$ is accepted, a new component with mean $\mu=0$ and the usual precision matrix will not make any difference. So, we make the following modifications:
\begin{itemize}
\item Allow the duplication of component means but there are no more than $n_{dup}$ duplicates at a specific location. ($x_{i_2}$ is a duplicate of $x_{i_1}$ if $x_{i_1}$, $x_{i_2}$ are close enough)
\item Assume that the component $N(\cdot|\mu=x_{i_1},Q=Q_{i_1})$ is added first at the location $x_{i_1}$. Then, more duplicates $x_{i_j}$ of $x_{i_1}$ are found and corresponding components $N(\cdot|\mu=x_{i_j},Q_{i_j})$ are added to the iterLap approximation. $Q_{i_j}$ is calculated by:
\begin{align}
\label{sec_miterLap_eqn:iterlap2_scalehess_02}
Q_{i_j} &= \kappa_{b}^{j} Q_{i_1},
\end{align}
with a user-defined factor $\kappa_b$
\end{itemize}
We call this multiplicative scaling which can be used in conjunction with the usual Hessian scaling. In that case, only the first Hessian $Q_{i_1}$ is scaled with $\kappa_a$. However, usually either Hessian scaling or multiplicative scaling is used. The approximation for Example \ref{sec_miterLap_exa:iterLap2_exa05} with multiplicative scaling ($n_{dup}=3$ and $\kappa_b=1.25$) is shown in Figure \ref{sec_miterLap_503_fig:ex05_approx02_dupmean_hessian_scale125}.

\vspace{+2em}
\noindent \textit{Other modifications}

\noindent There are some other modifications to iterLap, including:
\begin{itemize}
\item The functionality of manually adding optimisation starting points to $X_s$ and explored points to $X$. Generally, when a new normal component with mean $\mu$ is added to the iterLap approximation, aside from the starting points and explored points proposed by iterLap, a user can manually add points relative to the mean $\mu$. This is useful in the case a user knows something about the target density, e.g. the support or the variable correlation.  For example, a user may add more starting points, checking if there is any unexplored mode in the interested support. Another potential usage of this functionality is when $x$ can be decomposed into $(x_a,x_b)$ and the conditional expectation $E_{x_a|x_b}(x_a)$ is known by a function $f(x_b)$ (this frequently occurs in Bayesian inference with conjugate prior). Hence, when a new component with mean $\mu=(\mu_a,\mu_b)$ is added, it may be worth to check the locations $(f(\mu_b+\delta_b),\mu_b+\delta_b)$.
\item In the quadratic programming step for estimating $w$, we use weighted least squares:
\begin{align}
\label{sec_miterLap_eqn:iterlap2_weighted_least_squares}
w &= \argmin_{b} [ (y-Zb)^{T}(\omega_x I)(y-Zb) ], ~~~ b \geq 0.
\end{align}
where $\omega_{x;i}$ is a weight for a explored point $x_i$ in $X$. It is designed that a user can adjust the weight $\omega_{x;i}$ for a location of a component mean $\mu=x_i$. 
\item Make the code more robust to computer numerical issues, e.g. derive the matrix $Z$ from a normalised log version, make the Hessian matrix of $g_{b;n_c}(x)$ positive definite and scale the parameters in the quadratic programming step.
\end{itemize}

The iterLap version with all above modifications is named mod-iterLap and will be compared the original iterLap by some examples in the next section. In each example, both versions are run with a predefined maximum number of components $n_{c;max}$ but the resulting approximations may have less components than $n_{c;max}$ due to stopping rules. Furthermore, in mod-iterLap, we use a simplification step to remove insignificant components, e.g. components with normalised weights $w_{n;i}<e^{-5}$. This simplification reduces the computation cost when the approximation is used for other purposes.

\section{Comparison}
\label{sec:comp}

Firstly, we consider a density with a non-linear dependency in two-dimensional space. In this section, notations $\widetilde{q}_{o;x}(\cdot)$, $\widetilde{r}_{o;x}(\cdot)$ are the approximated densities of the original iterLap and $\widetilde{q}_{m;x}(\cdot)$, $\widetilde{r}_{m;x}(\cdot)$ are for the modified iterLap.

\begin{example}
\label{sec_comp_exa:iterLap2_exa06}
Define a target density $p_{x}(\cdot)$ on $x=(x_a,x_b)$:
\begin{align}
\label{sec_comp_eqn:iterLap2_exa06_targetq}
p_{x}(x) &= N(x_a|\mu=0,\sigma^2=10^2)N(x_b|\mu=0.03(x_a-3)^2+5,\sigma^2=1^2).
\end{align}
\end{example}

\begin{figure}[!ht]
\centering
\begin{adjustbox}{center}
\begin{subfigure}[b]{0.55\textwidth}
    \centering
    \includegraphics[scale=0.6]{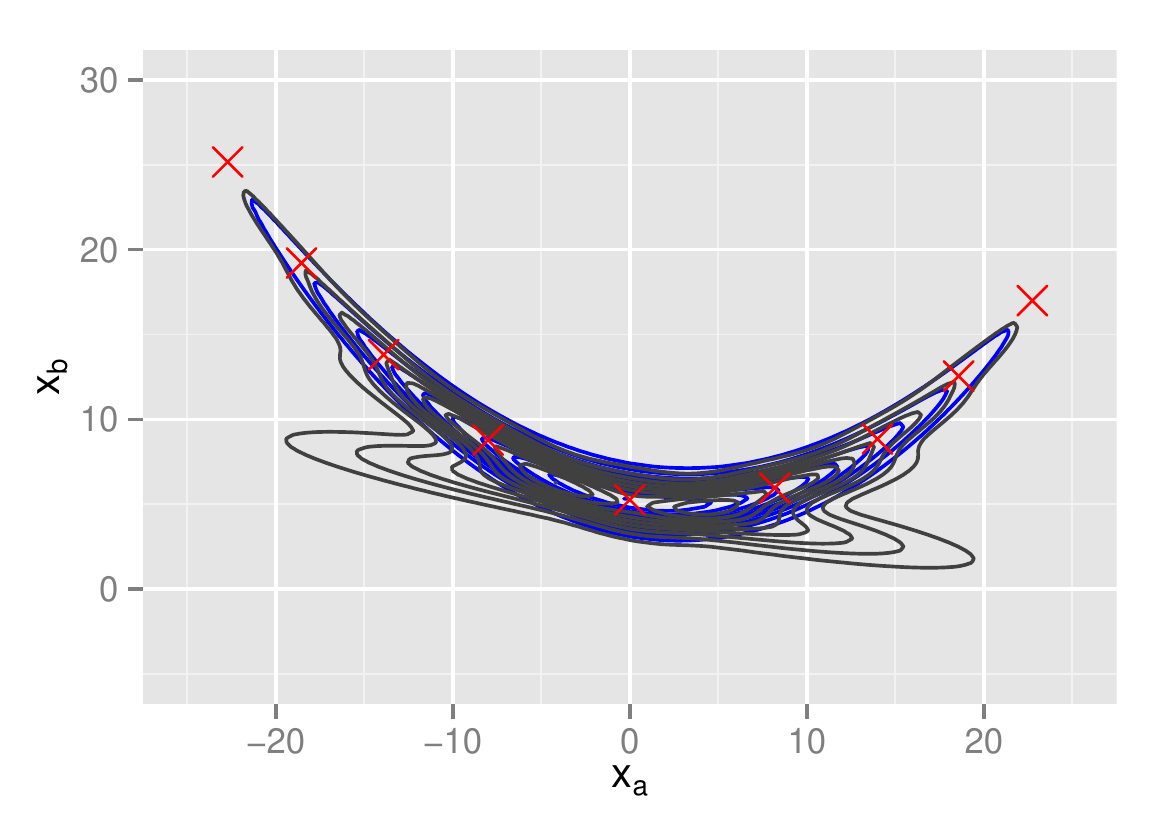}
    \subcaption{iterLap}
    \label{sec_comp_505_fig:ex06_contour_il01}
\end{subfigure}
\begin{subfigure}[b]{0.55\textwidth}
    \centering
    \includegraphics[scale=0.6]{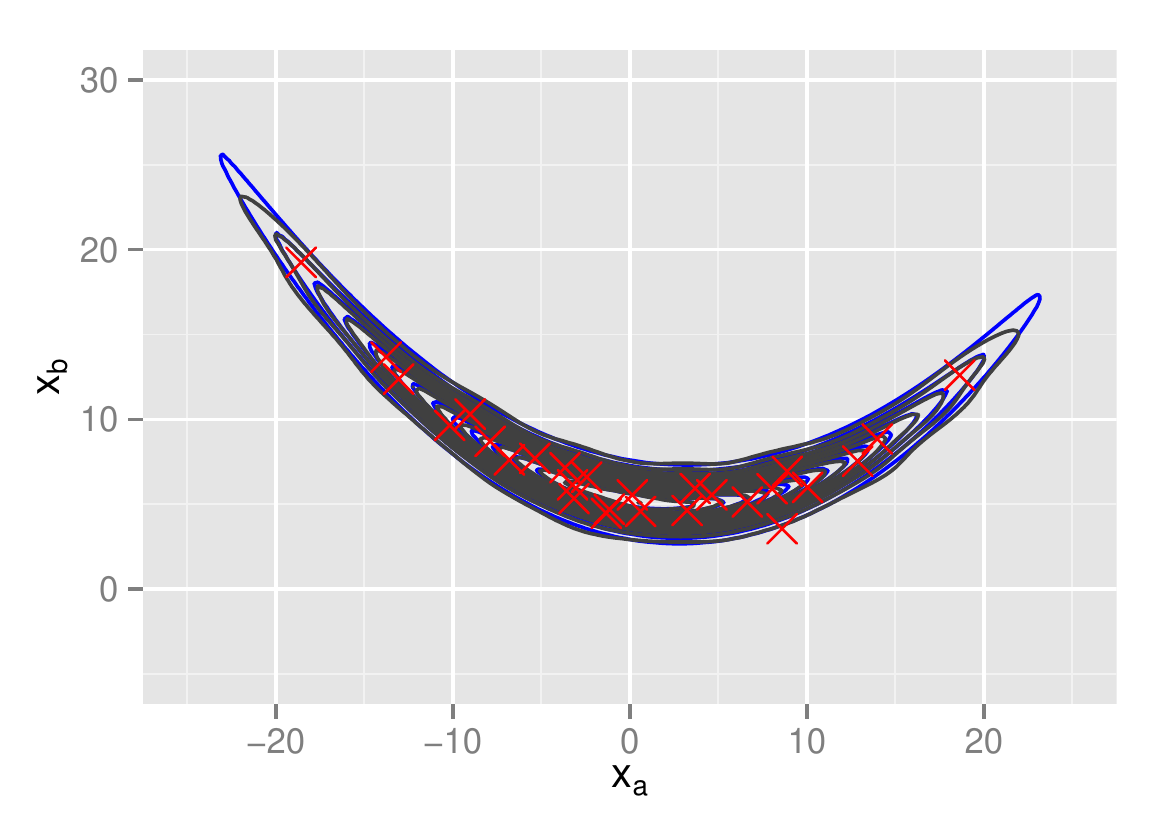}
    \subcaption{mod-iterLap}
    \label{sec_comp_505_fig:ex06_contour_il02}
\end{subfigure}
\end{adjustbox}

\caption{Example \ref{sec_comp_exa:iterLap2_exa06}: the blue and black contours are the target standardised density $r_{x}(\cdot)$ and the approximated standardised density $\widetilde{r}_{\cdot;x}(\cdot)$ respectively. The red crosses are the component means.}
\label{sec_comp_505_fig:ex06_contour}
\end{figure}

Both approximations are run with the maximum number of components $n_{c;max}=50$. $\widetilde{r}_{o;x}(\cdot)$ stops with $n_c=11$ components while $\widetilde{r}_{m;x}(\cdot)$ has $n_c=27$ components. The contours of the standardised densities are shown in Figure \ref{sec_comp_505_fig:ex06_contour}. Clearly, the modified version has a better capture of the non-linear dependency. 


\begin{figure}[!ht]
\centering
\begin{adjustbox}{center}
\begin{subfigure}[b]{0.55\textwidth}
    \centering
    \includegraphics[scale=0.6]{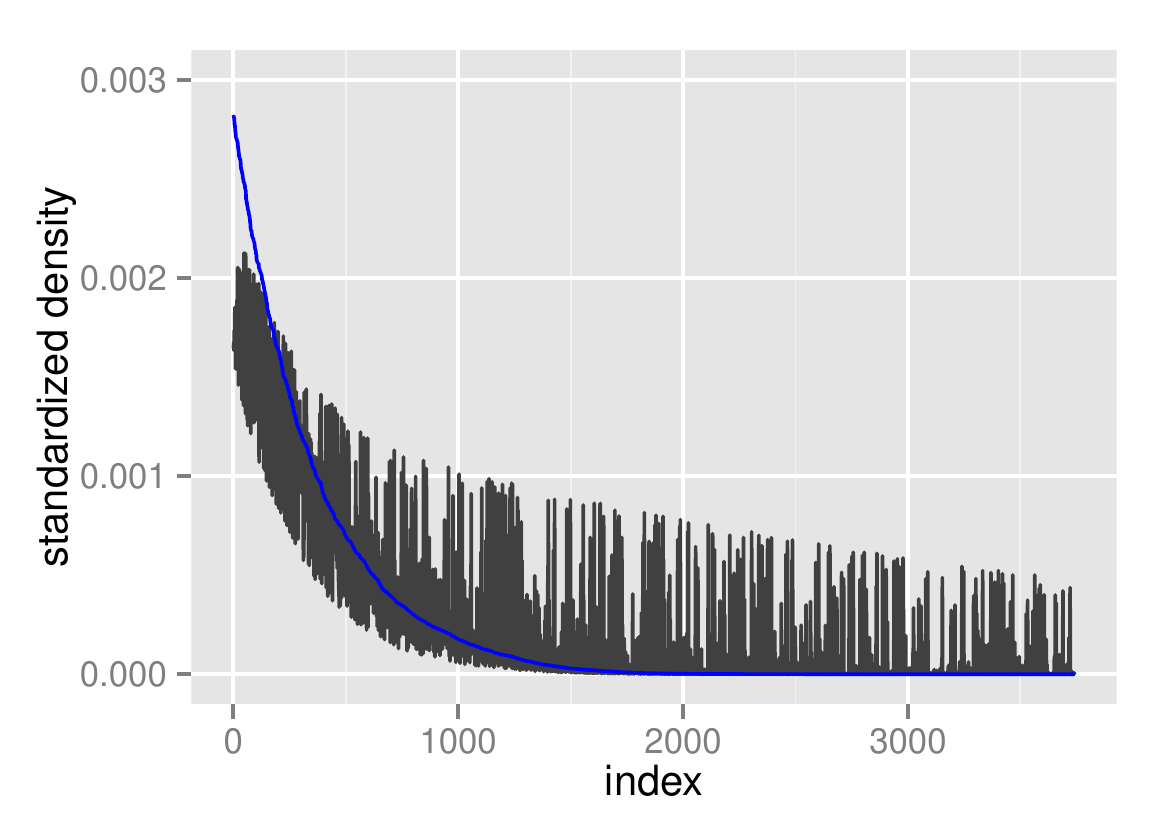}
    \subcaption{iterLap: $0.222$ seconds}
    \label{sec_comp_506_fig:ex06_ordered_stdden_il01}
\end{subfigure}
\begin{subfigure}[b]{0.55\textwidth}
    \centering
    \includegraphics[scale=0.6]{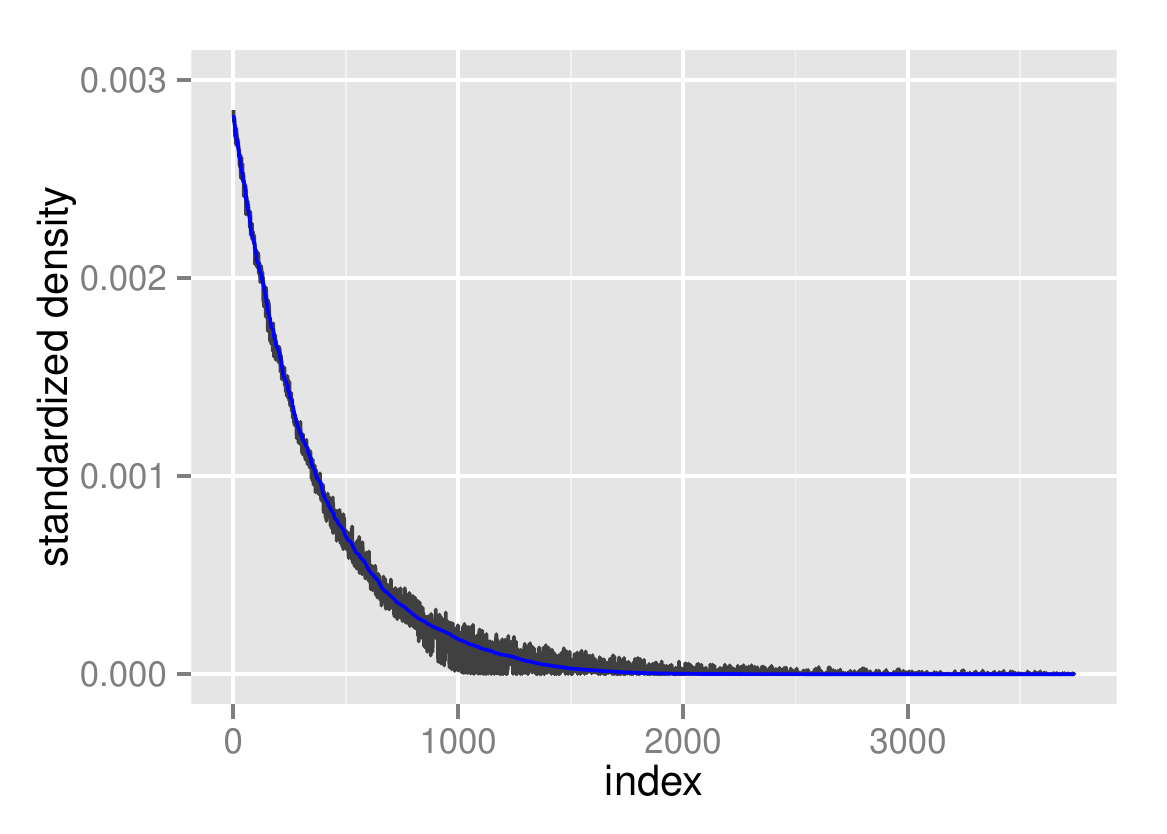}
    \subcaption{mod-iterLap: $1.142$ seconds}
    \label{sec_comp_506_fig:ex06_ordered_stdden_il02}
\end{subfigure}
\end{adjustbox}

\caption{Example \ref{sec_comp_exa:iterLap2_exa06}: the plots of the ordered standardised densities. The blue and black curves are $r_{x}(\cdot)$ and $\widetilde{r}_{\cdot;x}(\cdot)$ respectively.}
\label{sec_comp_506_fig:ex06_ordered_stdden}
\end{figure}

Figure \ref{sec_comp_505_fig:ex06_contour} shows the variable correlation but not the approximation error. So, to visually compare two approximations, $r_{x}(\cdot)$ is evaluated in a grid and then sorted in decreasing order of $r_{x}(\cdot)$. The approximated $\widetilde{r}_{\cdot;x}(\cdot)$ is sorted by the same order. Both of them are then plotted in Figure \ref{sec_comp_506_fig:ex06_ordered_stdden}. We also calculate the statistic of Equation \ref{sec_miterLap_eqn:iterlap2_standardised_density_compstat} for two approximations:  $s(r_{x},\widetilde{r}_{o;x})=0.424$ and  $s(r_{x},\widetilde{r}_{m;x})=0.078$. Running times in R language are included in the standardised density plots in this section.

The next example is for testing a density with non-linearity and multi-modality.

\begin{example}
\label{sec_comp_exa:iterLap2_exa07}
Define a target density $p_{x}(\cdot)$ on $x=(x_a,x_b)$:
\begin{align}
\label{sec_comp_eqn:iterLap2_exa07_targetq}
p_{x}(x) &= 0.5 N(x_a|\mu=-1,\sigma^2=6)N(x_b|\mu=-0.5(x_a+1)^2+3,\sigma^2=2)\\
\nonumber &~~~ + 0.5 N(x_a|\mu=1,\sigma^2=6)N(x_b|\mu=0.5(x_a-1)^2-3,\sigma^2=2).
\end{align}
\end{example}

\begin{figure}[!ht]
\centering
\begin{adjustbox}{center}
\begin{subfigure}[b]{0.55\textwidth}
    \centering
    \includegraphics[scale=0.6]{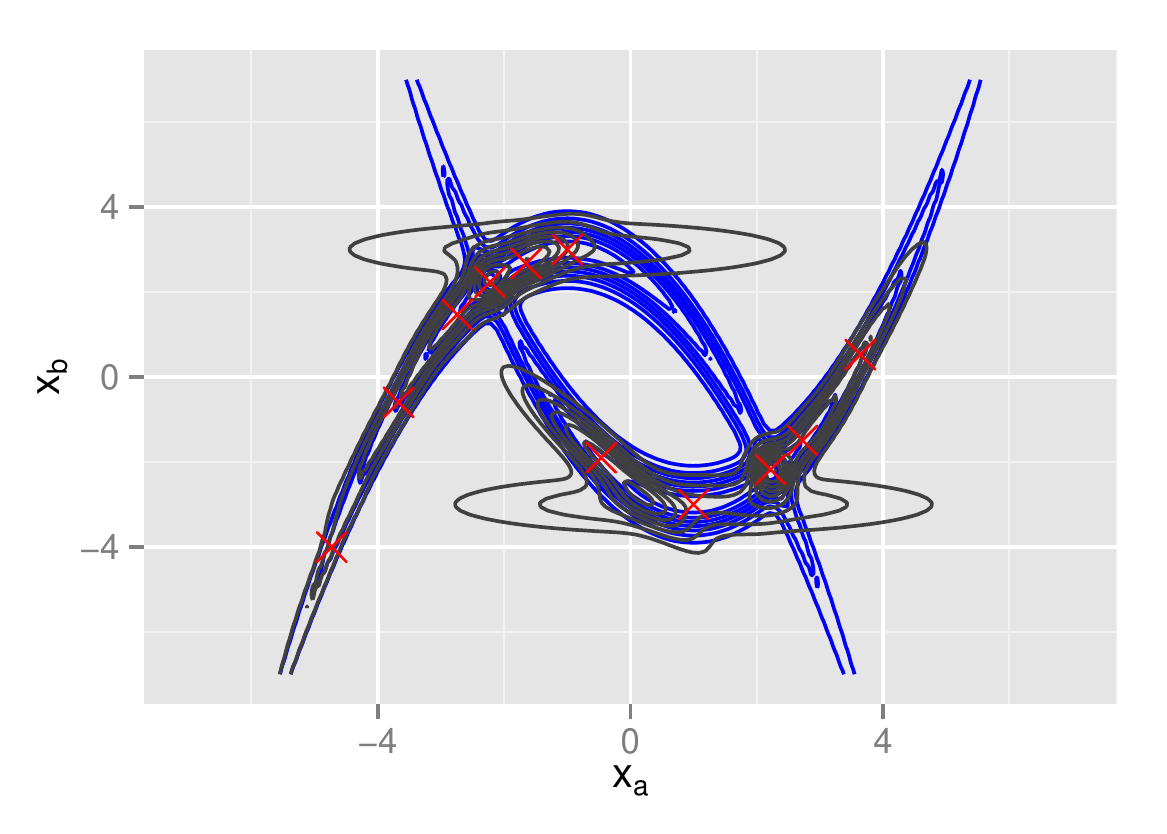}
    \subcaption{iterLap}
    \label{sec_comp_509_fig:ex07_contour_il01}
\end{subfigure}
\begin{subfigure}[b]{0.55\textwidth}
    \centering
    \includegraphics[scale=0.6]{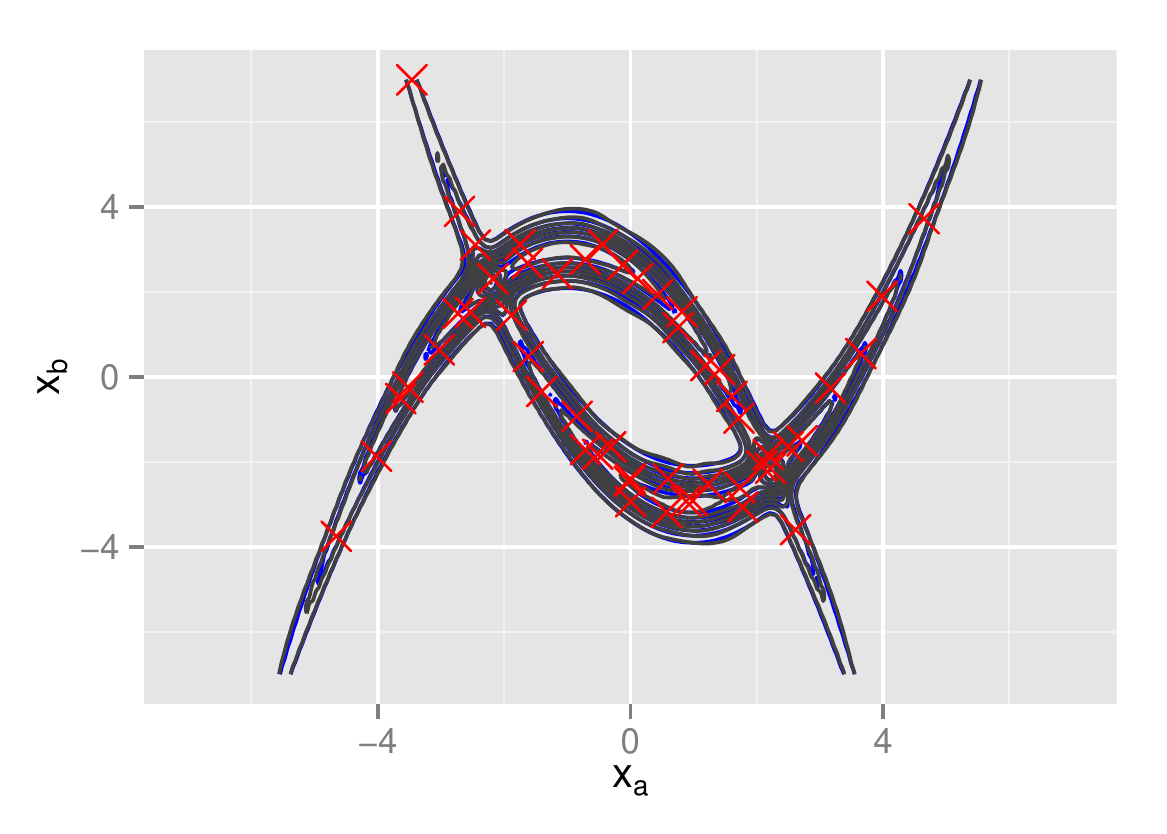}
    \subcaption{mod-iterLap}
    \label{sec_comp_509_fig:ex07_contour_il02}
\end{subfigure}
\end{adjustbox}

\caption{Example \ref{sec_comp_exa:iterLap2_exa07}: the blue and black contours are the target standardised density $r_{x}(\cdot)$ and the approximated standardised density $\widetilde{r}_{\cdot;x}(\cdot)$ respectively. The red crosses are the component means.}
\label{sec_comp_509_fig:ex07_contour}
\end{figure}

\begin{figure}[!ht]
\centering
\begin{adjustbox}{center}
\begin{subfigure}[b]{0.55\textwidth}
    \centering
    \includegraphics[scale=0.6]{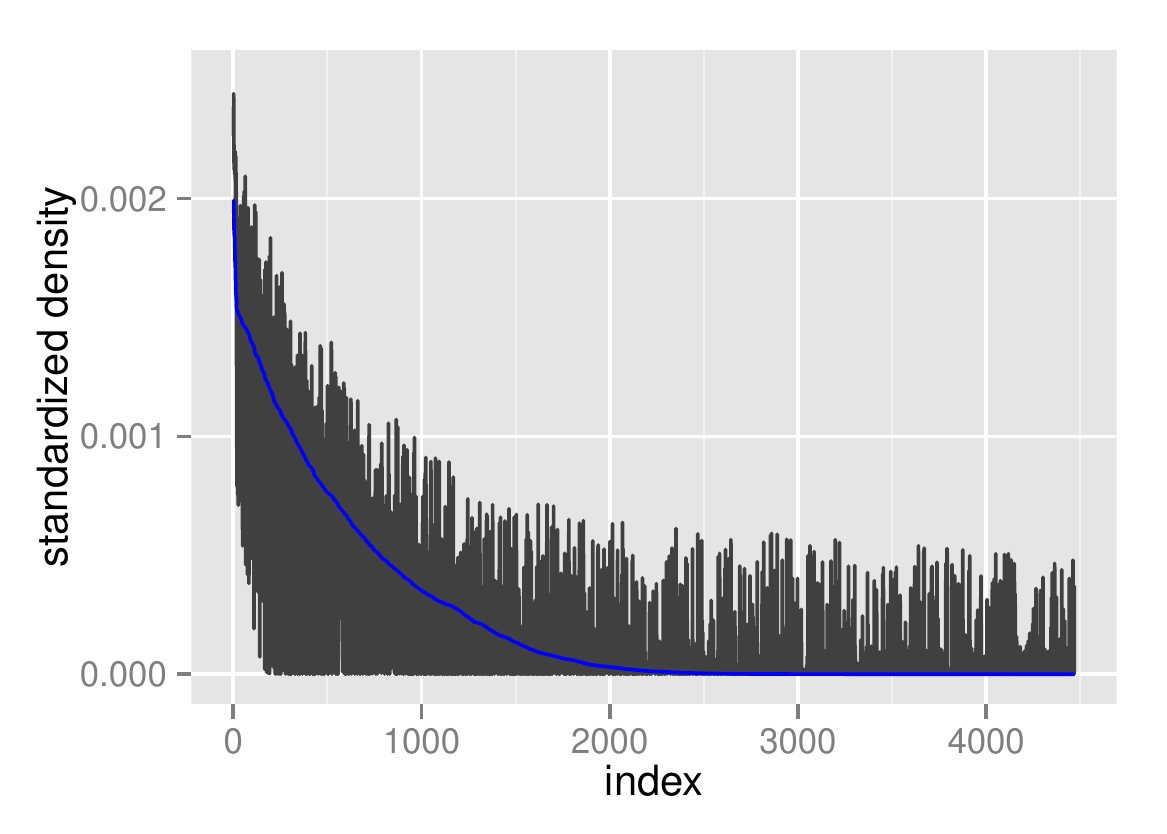}
    \subcaption{iterLap: $0.300$ seconds}
    \label{sec_comp_510_fig:ex07_ordered_stdden_il01}
\end{subfigure}
\begin{subfigure}[b]{0.55\textwidth}
    \centering
    \includegraphics[scale=0.6]{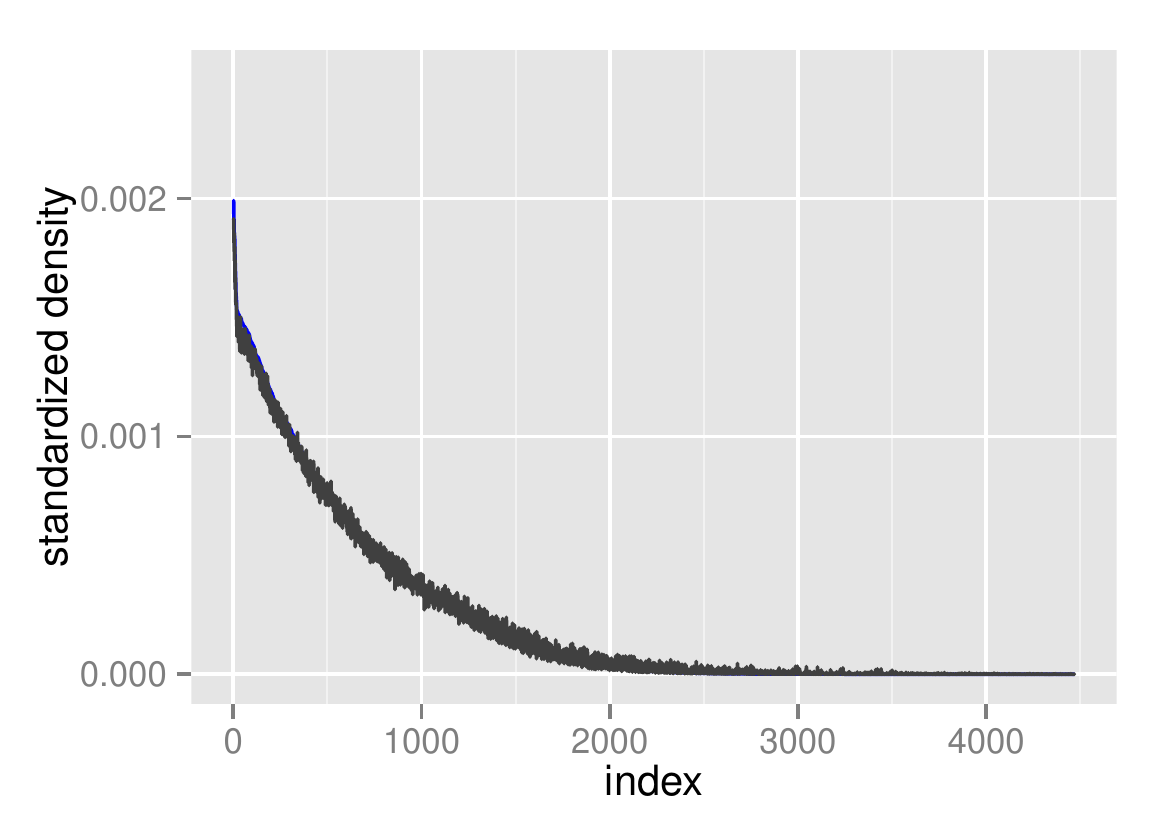}
    \subcaption{mod-iterLap: $5.090$ seconds}
    \label{sec_comp_510_fig:ex07_ordered_stdden_il02}
\end{subfigure}
\end{adjustbox}

\caption{Example \ref{sec_comp_exa:iterLap2_exa07}: the plots of the ordered standardised densities. The blue and black curves are $r_{x}(\cdot)$ and $\widetilde{r}_{\cdot;x}(\cdot)$ respectively.}
\label{sec_comp_510_fig:ex07_ordered_stdden}
\end{figure}

With $n_{c;max}=100$, iterLap stops with  $\widetilde{r}_{o;x}(\cdot)$ of $n_c=12$ components while mod-iterLap has $\widetilde{r}_{m;x}(\cdot)$ with $n_c=56$ components. The contour plots and ordered standardised density plots are shown in Figure \ref{sec_comp_509_fig:ex07_contour} and Figure \ref{sec_comp_510_fig:ex07_ordered_stdden}. $s(r_{x},\widetilde{r}_{o;x})=0.602$ and  $s(r_{x},\widetilde{r}_{m;x})=0.066$.


In Example \ref{sec_comp_exa:iterLap2_exa08}, we try increasing the dimension of the parameter space. 

\begin{figure}[!ht]
\centering
\begin{adjustbox}{center}
\begin{subfigure}[b]{0.55\textwidth}
    \centering
    \includegraphics[scale=0.6]{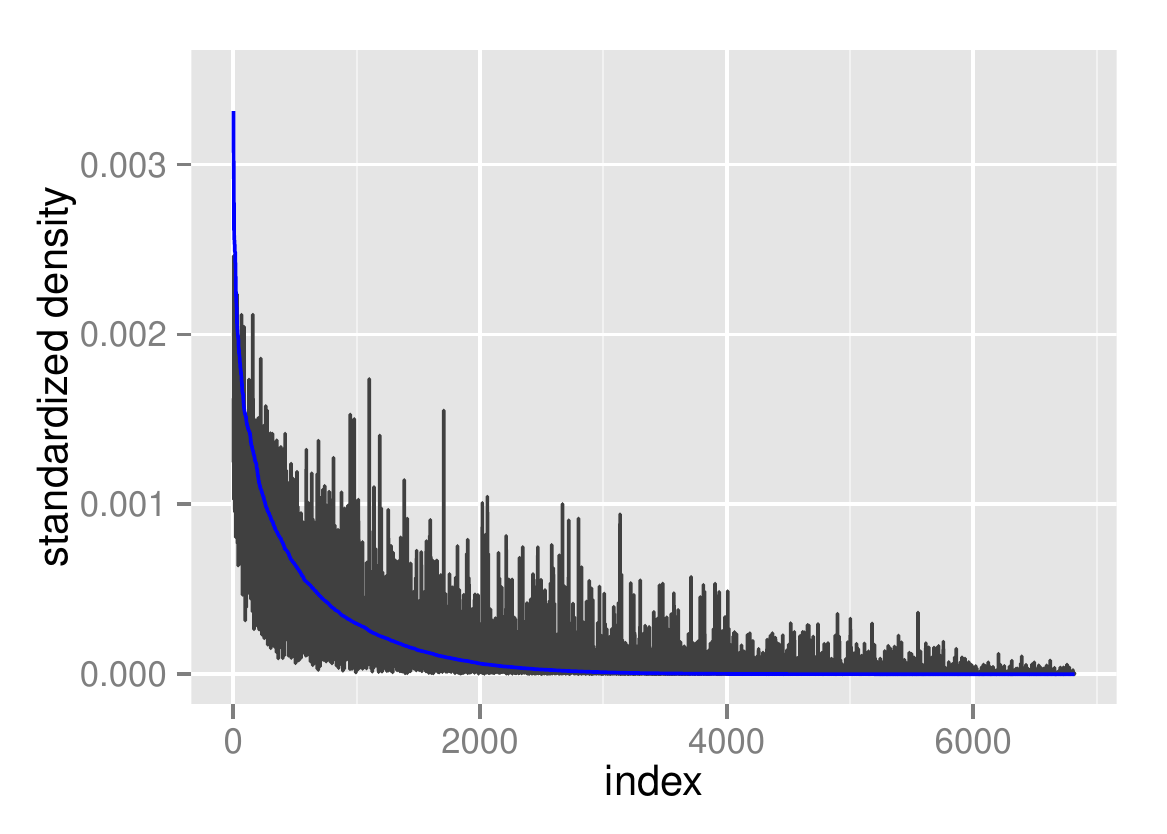}
    \subcaption{iterLap: $1.480$ seconds}
    \label{sec_comp_515_fig:ex08_ordered_stdden_il01}
\end{subfigure}
\begin{subfigure}[b]{0.55\textwidth}
    \centering
    \includegraphics[scale=0.6]{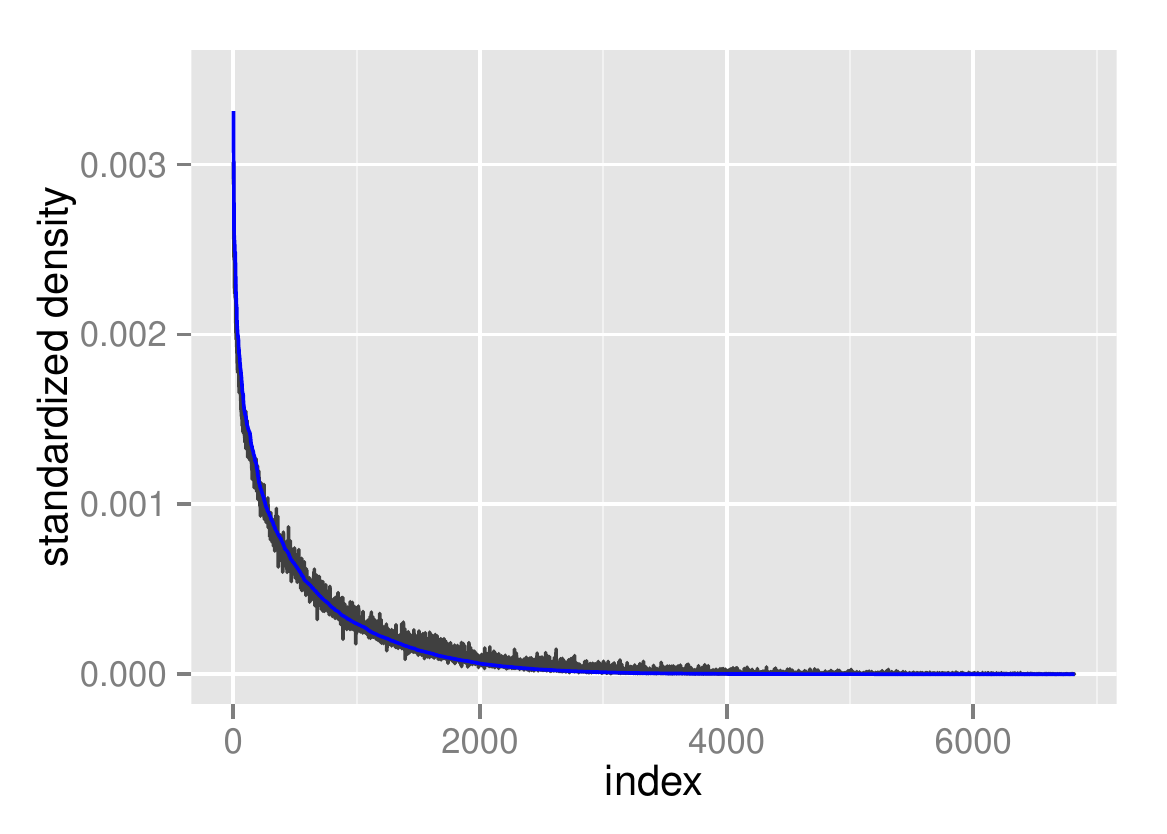}
    \subcaption{mod-iterLap: $122.934$ seconds}
    \label{sec_comp_515_fig:ex08_ordered_stdden_il01}
\end{subfigure}
\end{adjustbox}

\caption{Example \ref{sec_comp_exa:iterLap2_exa08}: the plots of the ordered standardised densities. The blue and black curves are $r_{x}(\cdot)$ and $\widetilde{r}_{\cdot;x}(\cdot)$ respectively.}
\label{sec_comp_515_fig:ex08_ordered_stdden}
\end{figure}

\begin{example}
\label{sec_comp_exa:iterLap2_exa08}
Define a target density $p_{x}(\cdot)$ on $x=(x_a,x_b,x_c)$ ($\dim(x_a)=\dim(x_b)=1$, $\dim(x_c)=4$):
\begin{align}
\label{sec_comp_eqn:iterLap2_exa08_targetq}
p_{x}(x) &= N(x_a|\mu_a,\Sigma_a=\sigma_a^2.I) \\
\nonumber &~~~~ N(x_b| \mu_b=A(x_a-\mu_a) + b,\Sigma_b=\sigma_b^2.I)\\
\nonumber &~~~~ N(x_c| \mu_c=C(x_1-\mu_a,x_b-\mu_b)^2,\Sigma_c=\sigma_c^2.I),
\end{align}
with:
\begin{alignat}{3}
\label{sec_comp_eqn:iterLap2_exa08_param}
\mu_a &=-0.5,~~~~~~~~~~~~~~~~&\sigma_a^2&=6.0,\\
\nonumber A &=-2.0,~~~~~~~~~~~~~~~~&b&=-1.0,\\
\nonumber \sigma_b^2&=0.2,~~~~~~~~~~~~~~~~&C&=\begin{pmatrix}
                        0.9    &  0.3\\
                        -0.3   &  -1.1\\
                        -0.5   &  -0.6\\
                        0.3    &  0.2
                        \end{pmatrix},\\
\nonumber \sigma_c^2&=(0.6,0.7,0.8,0.9)/3.
\end{alignat}
\end{example}
Notice that there is a linear dependency of $x_a$, $x_b$ and a non-linear dependency of $x_a$, $x_b$ and $x_c$. The conditional variances $\sigma_b^2$ and $\sigma_c^2$ define the dependency strength. As it is impossible to visualise this density, we only show the ordered standardised densities in Figure \ref{sec_comp_515_fig:ex08_ordered_stdden}. With $n_{c;max}=200$, $\widetilde{r}_{o;x}(\cdot)$ has $n_c=17$ components and $\widetilde{r}_{m;x}(\cdot)$ has $n_c=73$ components. $s(r_{x},\widetilde{r}_{o;x})=0.733$ and  $s(r_{x},\widetilde{r}_{m;x})=0.115$.


So, the functional approximation method still works with $\dim(x)=6$. We further increase the dimension to see its performance with Example \ref{sec_comp_exa:iterLap2_exa09}.

\begin{example}
\label{sec_comp_exa:iterLap2_exa09}
The same target density form in Equation \ref{sec_comp_eqn:iterLap2_exa08_targetq} is used with $\dim(x_a)=\dim(x_b)=2$, $\dim(x_c)=5$ and following parameters:
\begin{alignat}{3}
\label{sec_comp_eqn:iterLap2_exa08_param}
\mu_a &=(-0.5,-1.0),~~~~~~~~~~~~~~~~&\sigma_a^2&=(6.0,7.0),\\
\nonumber A &=\begin{pmatrix}
        0.5    &  -1.2\\
        -2.9  &  -1.3
        \end{pmatrix},~~~~~~~~~~~~~~~~&b&=(-1.0,-1.5),\\
\nonumber \sigma_b^2&=(0.2,0.3),~~~~~~~~~~~~~~~~&C&= \begin{pmatrix}
                        0.9    &  -1.3   &  -0.3   &  0.8\\
                        -0.7   &  0.8    &  -0.1   &  0.6\\  
                        0.7    &  -0.6   &  1.4    &  1.5\\
                        1.2    &  -1.2   &  0.3    &  0.0\\
                        1.3    &  1.4    &  1.4    &  0.0
                        \end{pmatrix},\\
\nonumber \sigma_c^2&=(0.8,0.9,1.0,1.1,1.2)/4.
\end{alignat}
\end{example}

\begin{figure}[!ht]
\centering
\begin{adjustbox}{center}
\begin{subfigure}[b]{0.55\textwidth}
    \centering
    \includegraphics[scale=0.6]{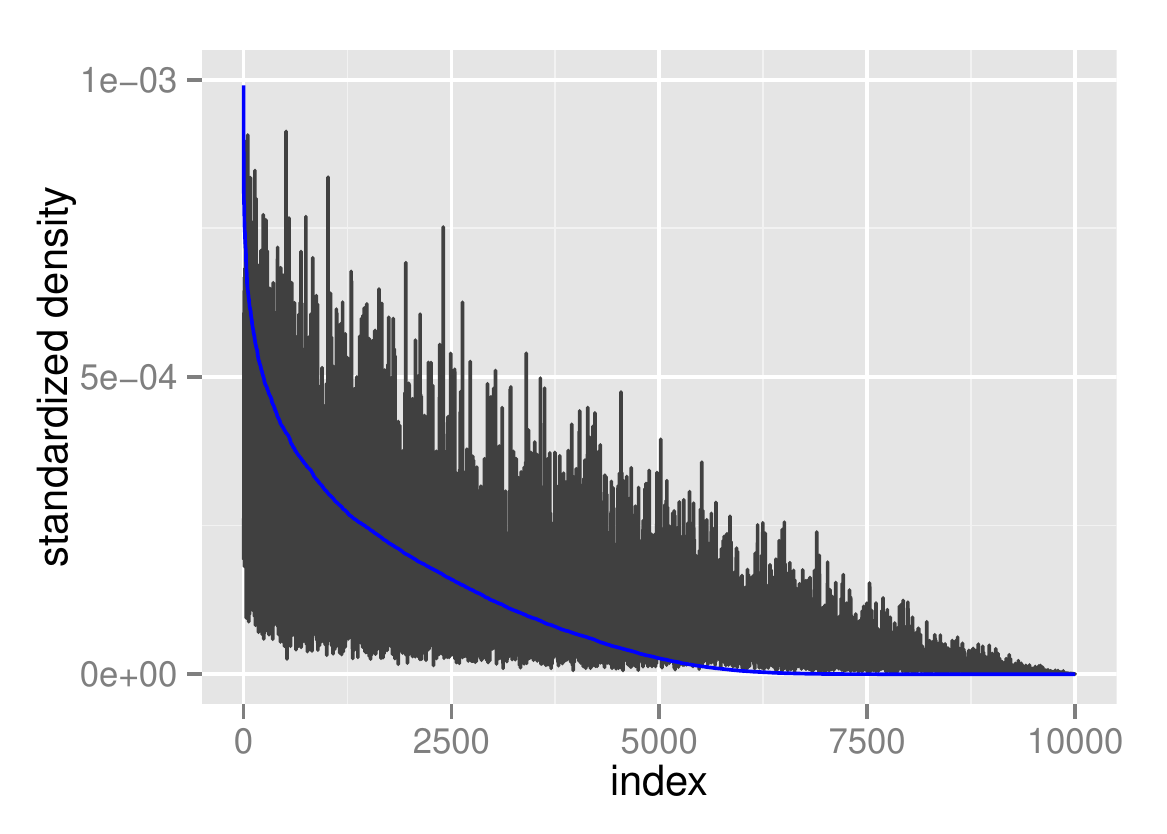}
    \subcaption{iterLap: $n_{c}=20$ ($4.441$ seconds)}
    \label{sec_comp_519_fig:ex09_ordered_stdden_il01}
\end{subfigure}
\begin{subfigure}[b]{0.55\textwidth}
    \centering
    \includegraphics[scale=0.6]{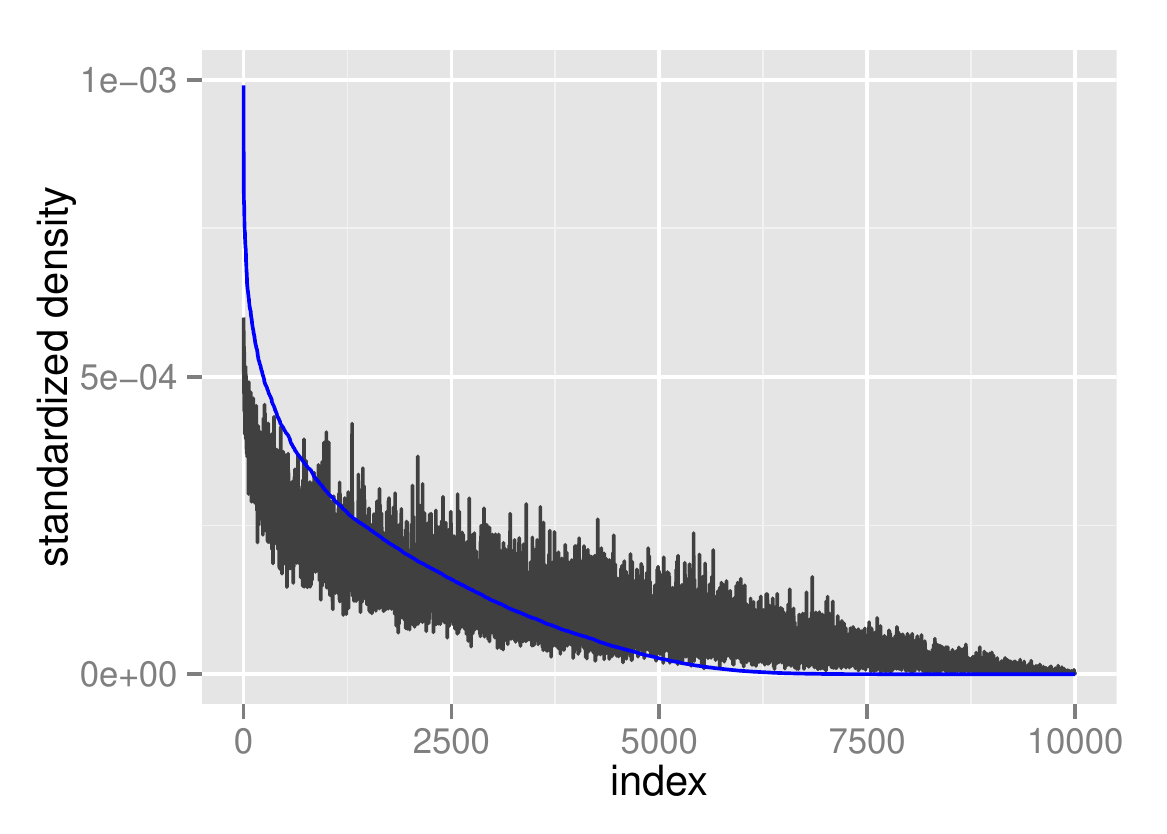}
    \subcaption{mod-iterLap: $n_{c}=50$ ($145.620$ seconds)}
    \label{sec_comp_519_fig:ex09_ordered_stdden_il02}
\end{subfigure}
\end{adjustbox}

\begin{adjustbox}{center}
\begin{subfigure}[b]{0.55\textwidth}
    \centering
    \includegraphics[scale=0.6]{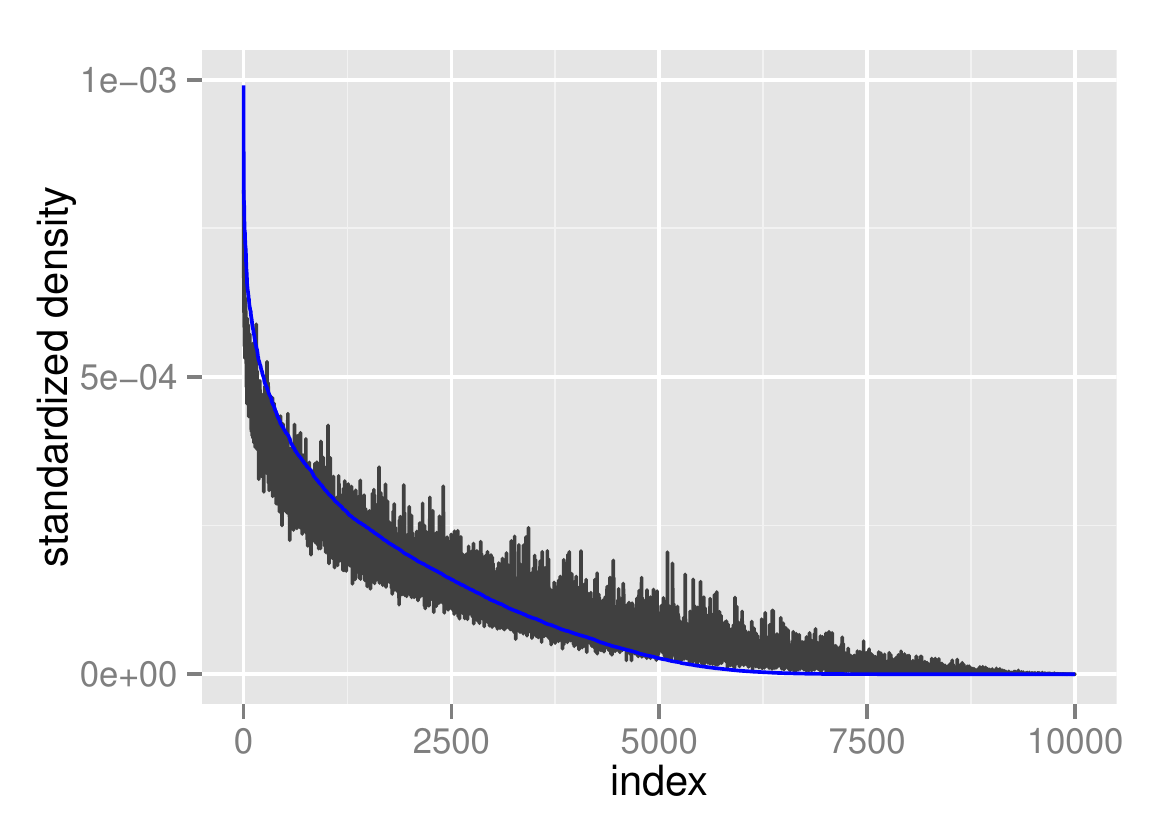}
    \subcaption{mod-iterLap: $n_{c}=237$ ($1.5$ hours)}
    \label{sec_comp_519_fig:ex09_ordered_stdden_il03}
\end{subfigure}
\begin{subfigure}[b]{0.55\textwidth}
    \centering
    \includegraphics[scale=0.6]{ch5_519_ex09_ordered_stdden_il03.pdf}
    \subcaption{mod-iterLap: $n_{c}=345$ ($7.0$ hours)}
    \label{sec_comp_519_fig:ex09_ordered_stdden_il04}
\end{subfigure}
\end{adjustbox}

\caption{Example \ref{sec_comp_exa:iterLap2_exa09}: the plots of the ordered standardised densities. The blue and black curves are $r_{x}(\cdot)$ and $\widetilde{r}_{\cdot;x}(\cdot)$ respectively.}
\label{sec_comp_519_fig:ex09_ordered_stdden}
\end{figure}

The iterLap and mod-iterLap approximations are first run with $n_{c;max}=200$. $\widetilde{r}_{o;x}(\cdot)$ and $\widetilde{r}_{m,a;x}(\cdot)$ have $n_{c}=20$ and $n_{c}=50$ components respectively and are plotted in Figures \ref{sec_comp_519_fig:ex09_ordered_stdden_il01} and \ref{sec_comp_519_fig:ex09_ordered_stdden_il02}. Even though $\widetilde{r}_{m,a;x}(\cdot)$ is better, it may not be very satisfying. Hence, we run mod-iterLap with more components and obtain $\widetilde{r}_{m,b;x}(\cdot)$ with $n_{c}=237$ components and $\widetilde{r}_{m,c;x}(\cdot)$ with $n_{c}=345$ components in Figures \ref{sec_comp_519_fig:ex09_ordered_stdden_il03} and \ref{sec_comp_519_fig:ex09_ordered_stdden_il04}. The approximations do get better with $s(r_{x},\widetilde{r}_{o;x})=0.763$,  $s(r_{x},\widetilde{r}_{m,a;x})=0.522$, $s(r_{x},\widetilde{r}_{m,b;x})=0.295$ and $s(r_{x},\widetilde{r}_{m,c;x})=0.244$ but approximation errors are not as good as ones of previous examples. 

So, like many other methods, iterLap suffers from the curse of dimensionality, especially when there is non-linear dependency between many variables.

In the last example, we will see how iterLap works with a non-normal noise and a constrained variable space. When used directly on such a constrained space, iterLap may have numerical issues in the optimisation and the Hessian evaluation at locations along the constrained border. Hence, it is better to transform a constrained space to a non-constrained space.

\begin{example}
\label{sec_comp_exa:iterLap2_exa10}
Consider a DLM:
\begin{alignat}{3}
\label{sec_comp_eqn:iterLap2_exa10_model}
y_t &= x_t + v_t &~~~~~ (t=1:n), \\
x_t &= x_{t-1} + u_t &~~~~~ (t=2:n), 
\end{alignat}
where $u_{t} \sim N(0,\sigma_u^2=\lambda_u^{-1})$, $v_{t} \sim N(0,\sigma_v^2=\lambda_v^{-1})$. The priors for precision parameters are $\lambda_u \sim Gamma(a=1,b=0.5)$, $\lambda_v \sim Gamma(c=1,d=0.5)$.
\end{example}

Notice that $x=x_{1:n}$ is an intrinsic Gaussian distribution with a precision matrix:
\begin{align}
\label{sec_comp_eqn:iterLap2_exa10_Qx}
Q_x &= \lambda_u \begin{pmatrix} 
          1  & -1 & 0  & 0  & \dots \\
          -1 & 2  & -1 & 0  & \dots \\
          0  & -1 & 2  & -1 & \dots \\
          \hdotsfor[2]{5}
          \end{pmatrix}.
\end{align}
One hundred data points $y_{1:n}$ $(n=100)$ are generated from Equation \ref{sec_comp_eqn:iterLap2_exa10_model} with $\lambda_u^{\star} = 1/2^2$, $\lambda_v^{\star} = 1/1^2$. The joint posterior of $(\lambda_u,\lambda_v,x)$ is:
\begin{align}
\label{sec_comp_eqn:iterLap2_exa10_joint_post}
p_{\lambda_u,\lambda_v,x|y}(\lambda_u,\lambda_v,x) &\propto Gamma(\lambda_u|a,b) Gamma(\lambda_v|c,d)\\
\nonumber &~~~~N(x|0,Q_x) N(y|x,Q_y=\lambda_v I_{n}).
\end{align}
It can be seen that the full conditional posterior of $x$ is:
\begin{align}
\label{sec_comp_eqn:iterLap2_exa10_fullcond_x}
p_{x|\lambda_u,\lambda_v,y}(x) = N(x|\mu_x^{\prime},Q_x^{\prime}),
\end{align}
with: 
\begin{align}
\label{sec_comp_eqn:iterLap2_exa10_fullcond_x_param}
Q_x^{\prime} &= Q_x + Q_y,\\
Q_x^{\prime} \mu_x^{\prime} &= Q_y y.
\end{align}
Using Equation \ref{sec_comp_eqn:iterLap2_exa10_fullcond_x} to marginalise Equation \ref{sec_comp_eqn:iterLap2_exa10_joint_post} with respect to $x$, we can obtain the marginal density $p_{\lambda_u,\lambda_v|y}(\lambda_u,\lambda_v)$:
\begin{align}
\label{sec_comp_eqn:iterLap2_exa10_marg_post_lambda}
p_{\lambda_u,\lambda_v|y}(\lambda_u,\lambda_v) &\propto q_{\lambda_u,\lambda_v|y}(\lambda_u,\lambda_v)\\
\nonumber &= Gamma(\lambda_u|a,b) Gamma(\lambda_v|c,d)\\
\nonumber &~~~~(2\pi)^{-(n-1)/2} \lambda_u^{(n-1)/2} |Q_x^{\prime}|^{-1/2} |Q_y|^{1/2}\\
\nonumber &~~~~\exp \left[ \frac{-y^T Q_y y + \mu_x^{\prime ~ T} Q_x^{\prime} \mu_x^{\prime} }{2}\right].
\end{align}

The log transform is used on $(\lambda_u,\lambda_v)$ to get $\tau_u = \log(\lambda_u)$ and $\tau_v = \log(\lambda_v)$, which are non-constrained variables. The corresponding non-normalised marginal density $q_{\tau_u,\tau_v|y}(\tau_u,\tau_v)$ is:
\begin{align}
\label{sec_comp_eqn:iterLap2_exa10_marg_post_tau}
q_{\tau_u,\tau_v|y}(\tau_u,\tau_v) &= q_{\lambda_u,\lambda_v|y}(\lambda_u,\lambda_v) \lambda_u \lambda_v,
\end{align}
which is then approximated by iterLap. Finally, the approximated density $\widetilde{q}_{\tau_u,\tau_v|y}(\tau_u,\tau_v)$ is converted back to $\widetilde{q}_{\lambda_u,\lambda_v|y}(\lambda_u,\lambda_v)$ by Equation \ref{sec_comp_eqn:iterLap2_exa10_marg_post_tau}.

\begin{figure}[!ht]
\centering
\begin{adjustbox}{center}
\begin{subfigure}[b]{0.55\textwidth}
    \centering
    \includegraphics[scale=0.6]{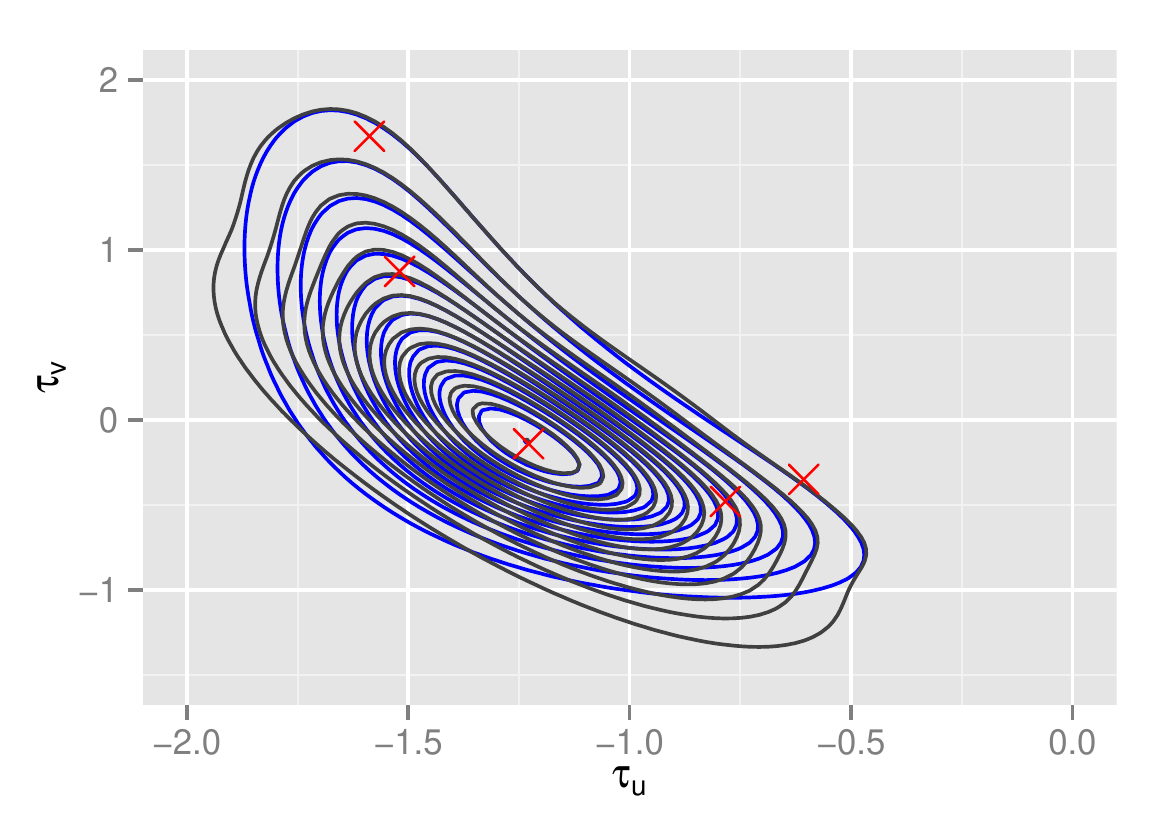}
    \subcaption{iterLap: $(\tau_u,\tau_v)$}
    \label{sec_comp_525_fig:ex10_contour_il01_tau}
\end{subfigure}
\begin{subfigure}[b]{0.55\textwidth}
    \centering
    \includegraphics[scale=0.6]{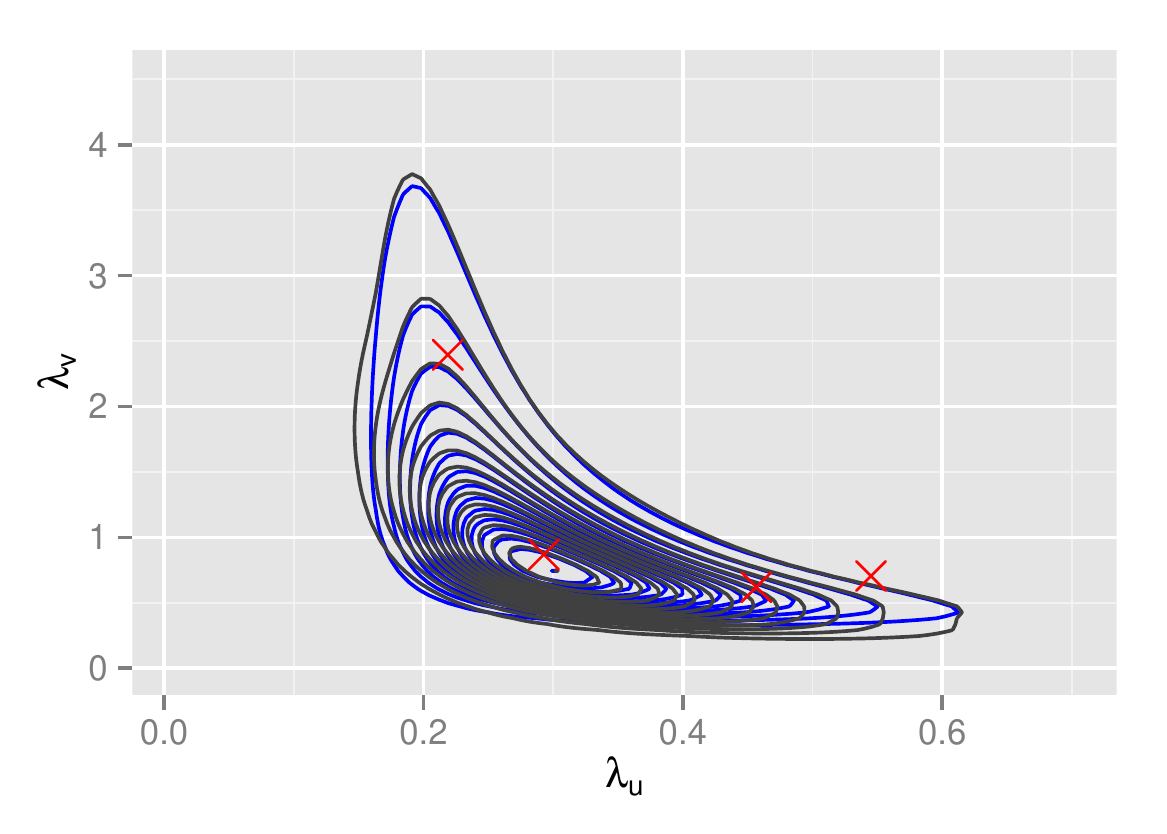}
    \subcaption{iterLap: $(\lambda_u,\lambda_v)$}
    \label{sec_comp_525_fig:ex10_contour_il01_lambda}
\end{subfigure}
\end{adjustbox}

\begin{adjustbox}{center}
\begin{subfigure}[b]{0.55\textwidth}
    \centering
    \includegraphics[scale=0.6]{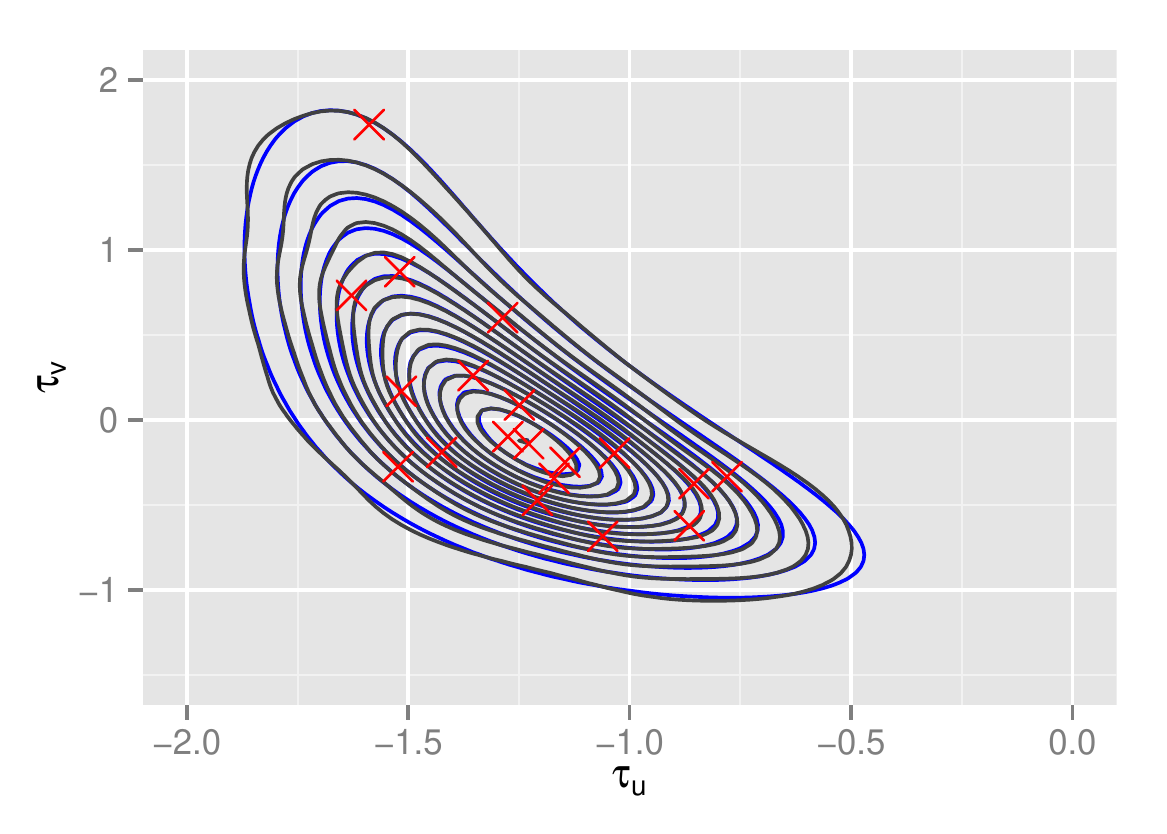}
    \subcaption{mod-iterLap: $(\tau_u,\tau_v)$}
    \label{sec_comp_525_fig:ex10_contour_il01_tau}
\end{subfigure}
\begin{subfigure}[b]{0.55\textwidth}
    \centering
    \includegraphics[scale=0.6]{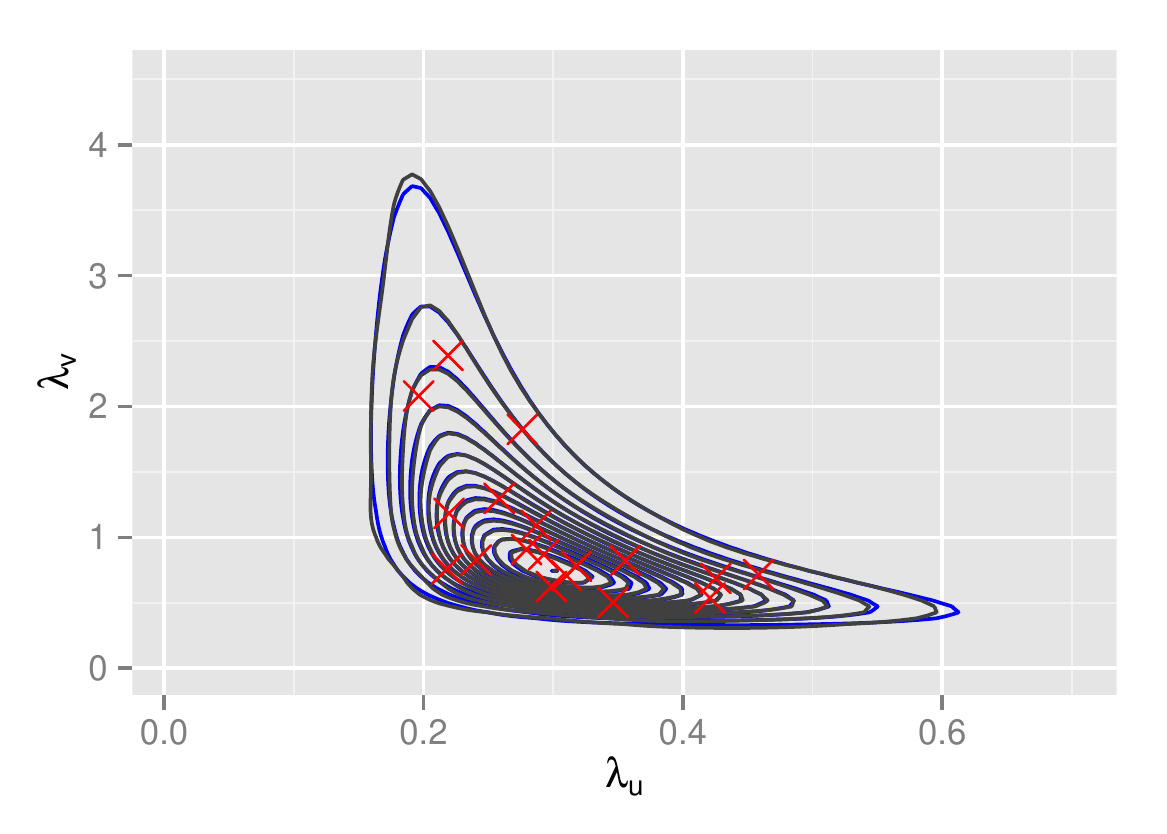}
    \subcaption{mod-iterLap: $(\lambda_u,\lambda_v)$}
    \label{sec_comp_525_fig:ex10_contour_il01_lambda}
\end{subfigure}
\end{adjustbox}

\caption{Example \ref{sec_comp_exa:iterLap2_exa10}: the blue and black contours are the target standardised density $r$ and the approximated standardised density $\widetilde{r}$ respectively. The red crosses are the iterLap component means (in the parametrisation $(\tau_u,\tau_v)$ or the corresponding transform $(\lambda_u,\lambda_v)$ of component means $(\tau_u,\tau_v)$. iterLap has $n_c=6$ components and mod-iterLap has $n_c=19$.}
\label{sec_comp_525_fig:ex10_contour}
\end{figure}

Two versions, iterLap and mod-iterLap, are run with $n_{c;max}=30$. The contour plots of standardised densities of $(\tau_u,\tau_v)$ and $(\lambda_u,\lambda_v)$ are shown in Figure \ref{sec_comp_525_fig:ex10_contour}

\begin{figure}[!ht]
\centering
\begin{adjustbox}{center}
\begin{subfigure}[b]{0.55\textwidth}
    \centering
    \includegraphics[scale=0.6]{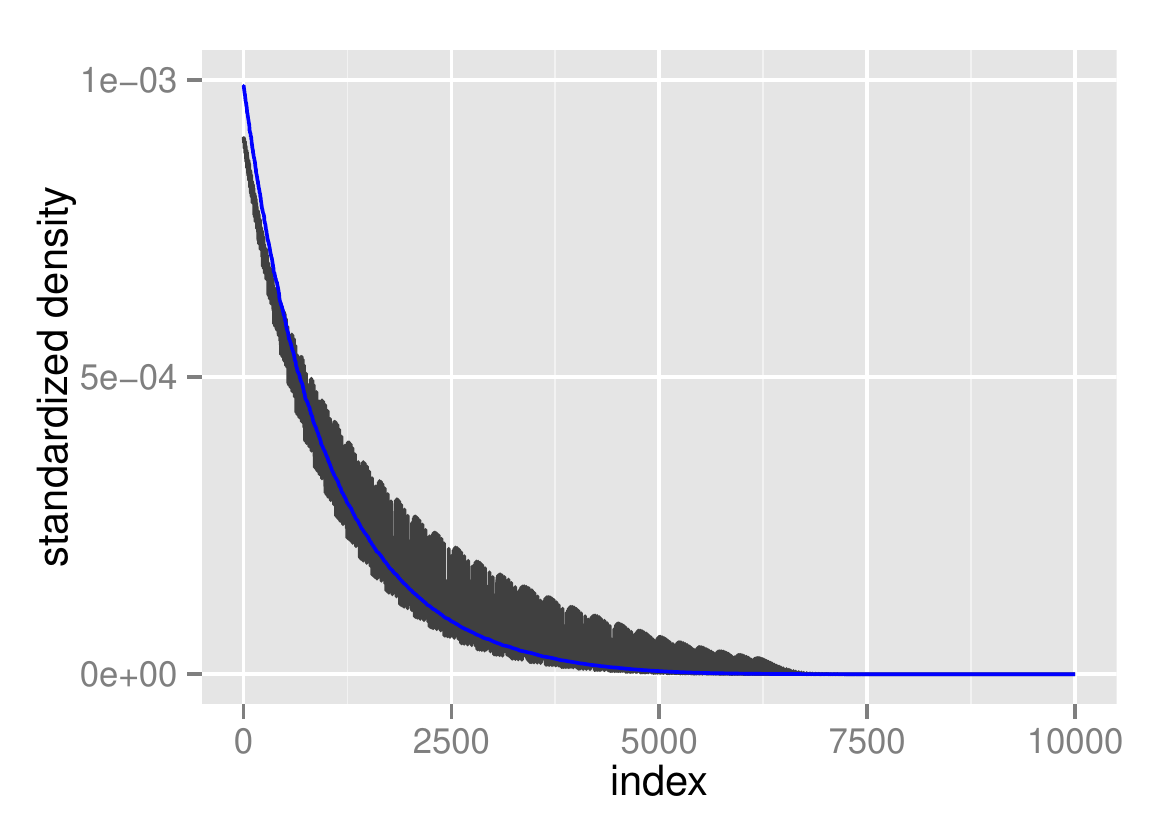}
    \subcaption{iterLap: $(\tau_u,\tau_v)$ ($0.498$ seconds)}
    \label{sec_comp_526_fig:ex10_ordered_stdden_il01_tau}
\end{subfigure}
\begin{subfigure}[b]{0.55\textwidth}
    \centering
    \includegraphics[scale=0.6]{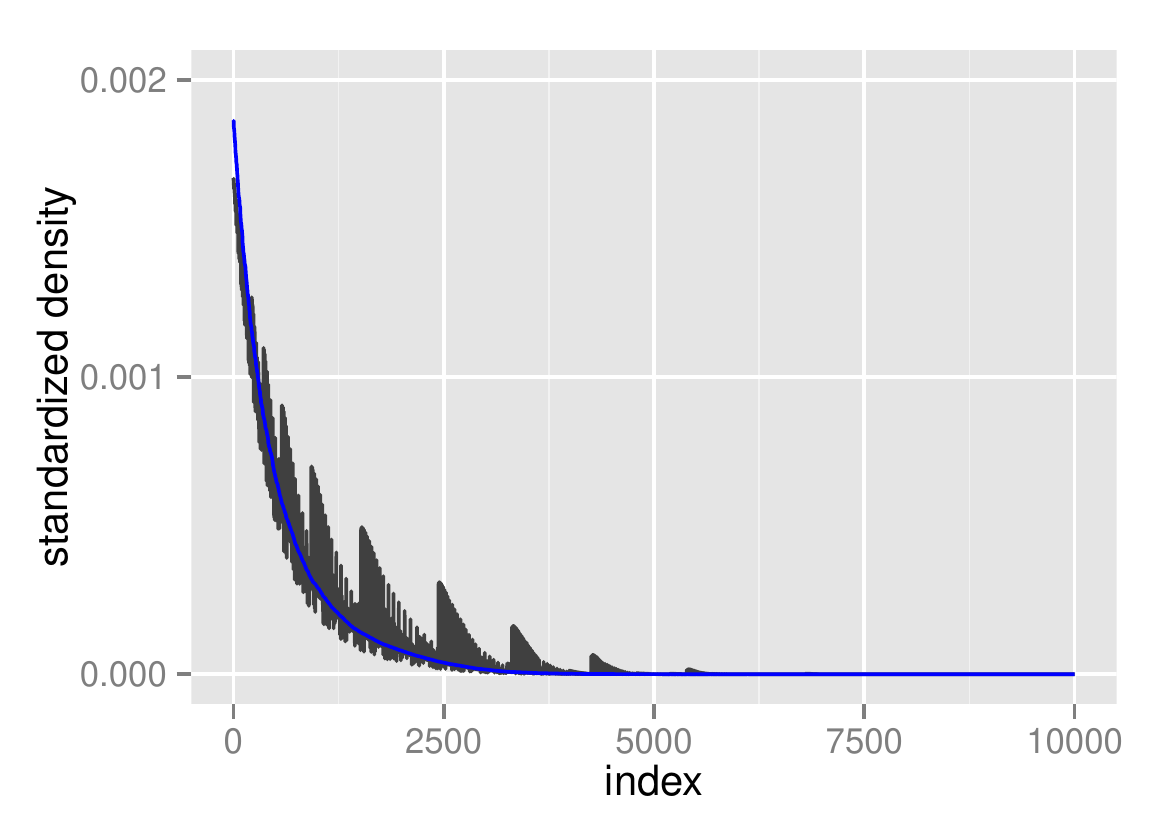}
    \subcaption{iterLap: $(\lambda_u,\lambda_v)$ ($0.498$ seconds)}
    \label{sec_comp_526_fig:ex10_ordered_stdden_il01_lambda}
\end{subfigure}
\end{adjustbox}

\begin{adjustbox}{center}
\begin{subfigure}[b]{0.55\textwidth}
    \centering
    \includegraphics[scale=0.6]{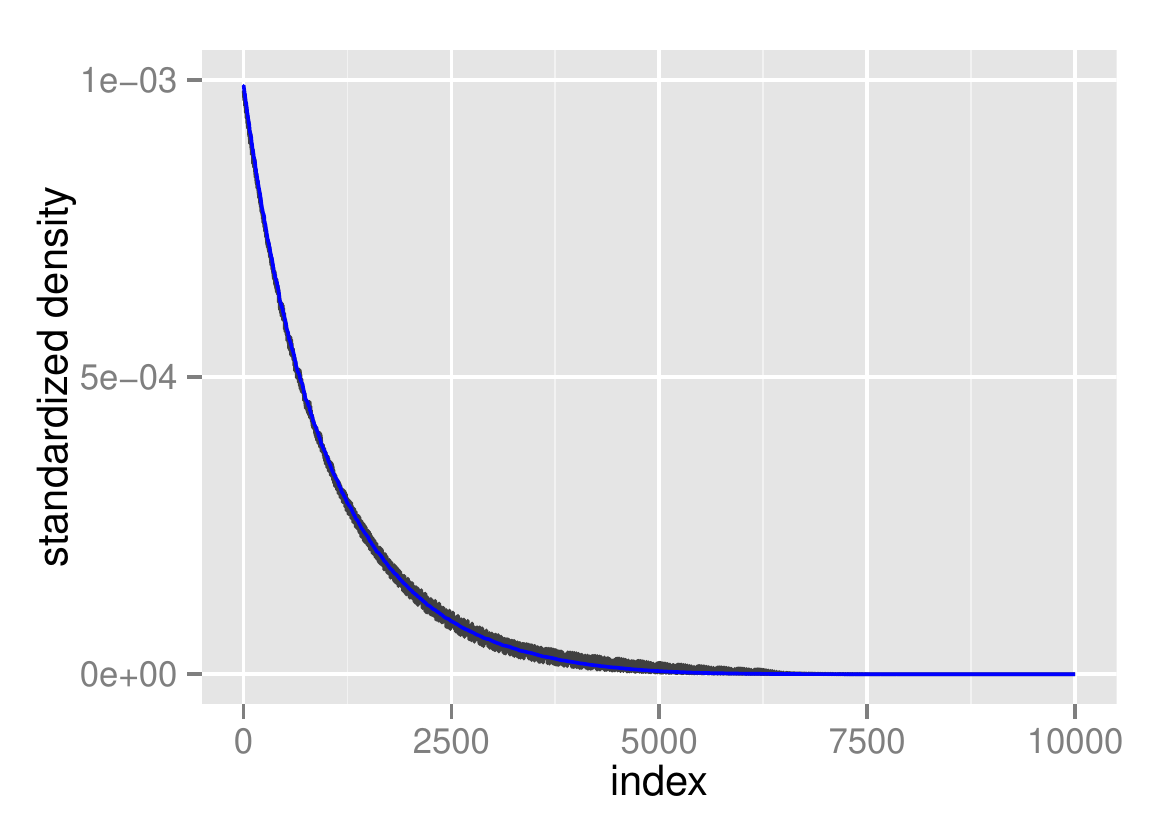}
    \subcaption{mod-iterLap: $(\tau_u,\tau_v)$ ($5.024$ seconds)}
    \label{sec_comp_526_fig:ex10_ordered_stdden_il02_tau}
\end{subfigure}
\begin{subfigure}[b]{0.55\textwidth}
    \centering
    \includegraphics[scale=0.6]{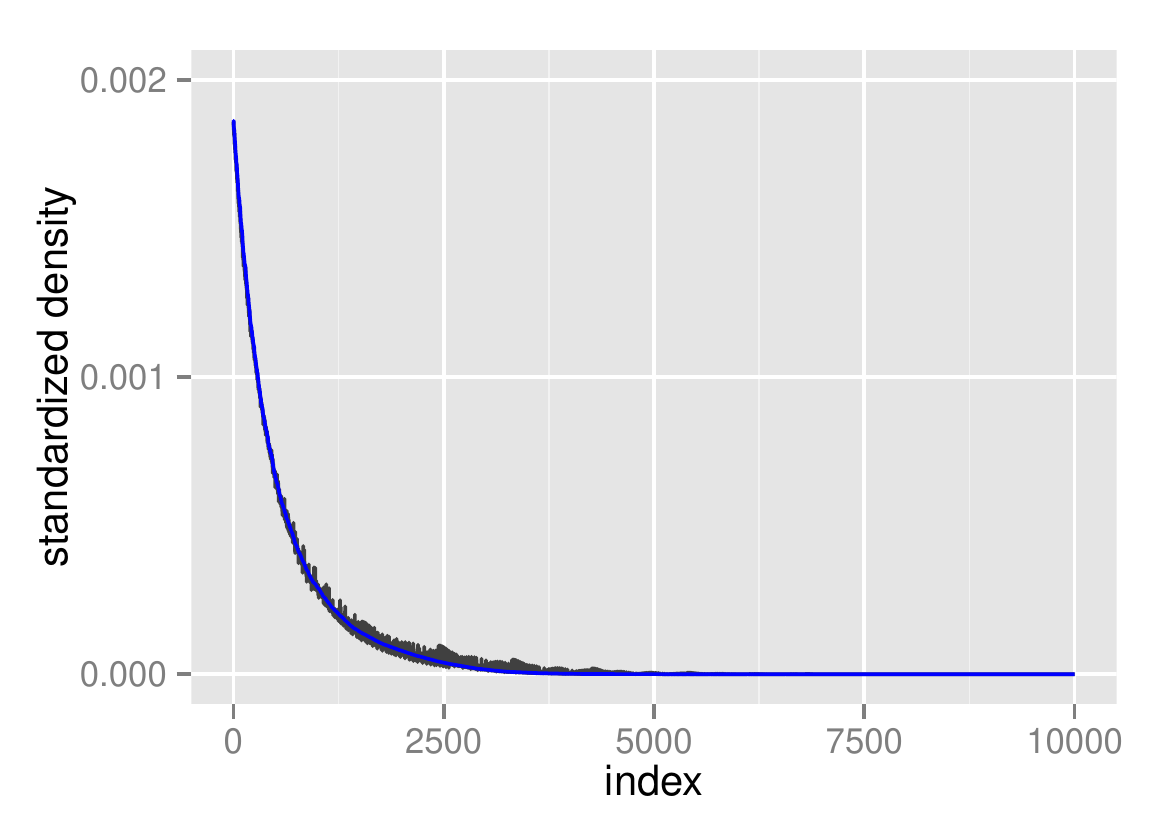}
    \subcaption{mod-iterLap: $(\lambda_u,\lambda_v)$ ($5.024$ seconds)}
    \label{sec_comp_526_fig:ex10_ordered_stdden_il02_lambda}
\end{subfigure}
\end{adjustbox}

\caption{Example \ref{sec_comp_exa:iterLap2_exa10}: the plots of the ordered standardised densities. The blue and black curves are $r_{x}(\cdot)$ and $\widetilde{r}_{\cdot;x}(\cdot)$ respectively.}
\label{sec_comp_526_fig:ex10_ordered_stdden}
\end{figure}

The ordered standardised densities are plotted in Figure \ref{sec_comp_526_fig:ex10_ordered_stdden}. In the parametrisation $(\tau_u,\tau_v)$, $s(r,\widetilde{r}_{o})=0.125$, $s(r,\widetilde{r}_{m})=0.03$ while in the parametrisation $(\lambda_u,\lambda_v)$, $s(r,\widetilde{r}_{o})=0.135$, $s(r,\widetilde{r}_{m})=0.032$.

Instead of transforming the densities back and forth between constrained space and non-constrained space, which involves the evaluation of a Jacobian, it may be more practical to work directly on the non-constrained space in some cases, e.g. specify the prior and approximate the posterior on $(\tau_u,\tau_v)$ in Example \ref{sec_comp_exa:iterLap2_exa10}. 

\section{Conclusion}
\label{sec:conclusion}

We have proposed a new solution to iterLap approximation with various implementation modifications such as stopping rule adjustment, proposal of new residual function, starting point selection for numerical optimisation, scaling of Hessian matrix. In all examples of Section \ref{sec:comp}, mod-iterLap achieves better performance with longer running time. This computation cost is reasonable as the mod-iterLap add more components to correct the approximation without getting stuck like the original iterLap. The more the number of components, the longer the running time. In practice, the trade-off between correctness and running time can be controlled by the maximum number of components $n_{c;max}$. Another point is that the code of all these examples, iterLap, mod-iterLap is written in R. Hence, the running time should improve significantly if the code is ported to C language. 

Such a functional approximation provides a fast approximation to any target density without relying on sampling which is another difficult and complex problem. For such sampling problems, iterLap can be used as a non-linear multi-modal sampling proposal in Monte Carlo methods, providing an efficient way to explore parameter space \citep{mref:Mai_PhdThesis2013}. This is also the direction our future work.

\bibliographystyle{splncs03}
\bibliography{mod_iterLap_nocref}

\end{document}